\documentclass{LMCS}

\usepackage{amssymb}
\usepackage{graphicx}

\usepackage{url}

\usepackage{amsmath,bbm}
\usepackage{enumerate}
\usepackage{qsymbols}
\usepackage{tabularx}
\usepackage{bussproofs}
\usepackage{mathpartir}
\usepackage{wasysym}
\usepackage{hyperref}

\newcommand{\lc}{\ensuremath{\lambda_{\mathcal{C}}}} 
\newcommand{\lcm}{\ensuremath{`l^-_{\mathcal{C}}}} 
\newcommand{\cterms}{\ensuremath{`L_0}} 
\newcommand{\cas}[2]{\{ \mskip-4mu | #1 | \mskip-4mu \}\cdot #2} 
\newcommand{\subs}[2]{[#1 := #2]}         
\newcommand{\val}{\ensuremath{\, \mathcal{V}\,}} 
\newcommand{\va}[1]{\ensuremath{\mathcal{V}al(#1)}} 
\newcommand{\fv}[1]{\mathcal{F\!V}(#1)}   
\newcommand{\red}[2]{\ensuremath{Red_{#1}(#2 )}} 

\newcommand{\these}{\vdash}               
\newcommand{\valid}{\vDash}               
\newcommand{\tv}[1]{\mathcal{T\!V}(#1)}   
\newcommand{\clo}[1]{\ensuremath{\overline{#1}}}  
\newcommand{\cre}{\ensuremath{\mathcal C\!R}} 
\newcommand{\dc}{\ensuremath{\mathcal D\!C}} 
\newcommand{\pcr}{\ensuremath{\mathcal P\!C\!R} } 
\newcommand{\cri}{\ensuremath{(\mathbf{C\!R1})}} 
\newcommand{\crd}{\ensuremath{(\mathbf{C\!R2})}} 
\newcommand{\crdb}{\ensuremath{(\mathbf{C\!R2'})}} 
\newcommand{\crt}{\ensuremath{(\mathbf{C\!R3})}} 
\newcommand{\crq}{\ensuremath{(\mathbf{C\!R4})}} 
\newcommand{\crqb}{\ensuremath{(\mathbf{C\!R4'})}} 
\newcommand{\cro}{\ensuremath{(\mathbf{C\!R})}} 
\newcommand{\itp}[2]{[#1]_{#2}}   
\newcommand{\itpcb}[2]{\llbracket #1 \rrbracket_{#2}} 
\newcommand{\acc}[3]{ #1 `: \itp{#2}{#3} } 

\newcommand{\pn}{\ensuremath{P\!N_0}} 
\newcommand{\neu}{\ensuremath{\mathcal{N}_D}} 

\newcommand{\cred}[1][]{\ensuremath{\xrightarrow[]{#1}_{\text{c}}}}
\newcommand{\ccalc}{\ensuremath{`l_{\mathtt{com}}}}
\newcommand{\bred}[1][]{\ensuremath{\to_{\scriptscriptstyle \mathcal{B}}}}
\newcommand{\bcalc}{\ensuremath{`l_{\mathcal{B}}}}

\newcommand{\co}[1]{\texttt{#1}}  
\newcommand{\tco}[1]{\textbf{#1}}  

\newcommand{\sw}[2][]{\ensuremath{\downarrow \!{#2}_{#1}}} 
\newcommand{\mes}[1]{\ensuremath{s(#1)}} 

\newcommand{\rul}[1]{\ensuremath{\textsc{\footnotesize{#1}}}} 
\newcommand{\trul}[1]{\ensuremath{\mathbf{#1}}} 
\newcommand{\ttrul}[1]{\ensuremath{\mathbf{\scriptstyle{#1}}}} 
\newcommand{\strul}[1]{\ensuremath{\mathtt{\scriptstyle{#1}}}} 
\newcommand{\srul}[1]{\ensuremath{\mathtt{#1}}} 
\newcommand{\nat}{\ensuremath{\mathit{n\!a\!t}}} 
\newcommand{\ctab}{\ensuremath{\co{c}_{\diamond}}} 
\newcommand{\ttab}{\ensuremath{\tco{c}_{\diamond}}} 

\newqsymbol{`?}{\maltese}
\newqsymbol{`<}{\preccurlyeq}
\newqsymbol{`0}{[\,]} 

\renewcommand{\vec}[1]{\overrightarrow{#1}}

\def\doi{7 (1:2) 2010}
\lmcsheading%
{\doi}
{1--24}
{}
{}
{Nov.~22, 2009}
{Mar.~16, 2011}
{}

\begin{document}


\title[Semantics of Typed Lambda-Calculus
  with Constructors]{Semantics of Typed Lambda-Calculus\\
  with Constructors}
\author[B.~Petit]{Barbara Petit}
\address{LIP - ENS Lyon\\
  46 All\'ee d'Italie, 69364 Lyon, France}
\urladdr{http://perso.ens-lyon.fr/barbara.petit}


\maketitle

\begin{abstract}
We present a Curry-style second-order type system with union and
intersection types for the lambda-calculus with constructors of
Arbiser, Miquel and Rios, an extension of lambda-calculus with a
pattern matching mechanism for variadic constructors. 
We then prove the strong normalisation and the absence of match
failure for a restriction of this
system, by adapting the standard reducibility method. 
\end{abstract}

\section*{Introduction}

Pattern matching is a crucial feature in modern programming
languages.
It appeared in the late 60's~\cite{snobol}, first as a simple
detection of rigidly specified values.
Although it still has this basic form in most imperative languages
(as the \texttt{case} of Pascal or the \texttt{switch} of C),
it now comes with more elaborated features in main functional
programming languages~\cite{Sml,Haskell,Ocaml} and proof assistants
(especially those based on type theory~\cite{Coq,Agda}).
In particular, the pattern matching ``\`a la ML'' is able to decompose
complex data-structures.

From the theoretical point of view, many approaches have been proposed
to extend lambda-calculus~\cite{Bar84} with pattern matching facilities,
such as the \emph{Rho-calculus}~\cite{CirKir98}, the
\emph{Pure pattern calculus}~\cite{JayKes06} and the
\emph{Lambda calculus with constructors}~\cite{alex06}.
Typed versions have also been presented for such
calculi~\cite{CirKir03,Jay04,Petit09,JayBook}.

The lambda-calculus with constructors~\cite{AAAjournal} decomposes
the pattern matching \textit{\`a la ML} using a case construct
$$\cas{c_1 \mapsto u_1; \ldots ; c_n \mapsto u_n}{t}$$
performing case analysis on constant constructors, in the spirit of the
\texttt{case} of Pascal.
Composite data structures consist of constructor applied to one or many
arguments.
Their destruction is achieved using a commutation
rule between case and application\footnotemark:
\footnotetext{Which differs from the commutative conversion
  rules~\cite{girard89} coming from logic.}
$$(\rul{CaseApp}) \qquad \qquad \cas{`q}{(tu) \;=\; (\cas{`q}{t})u }$$
Thanks to this rule, one can encode the whole ML-style pattern
matching in the calculus, and write destruction functions on more
complex data types, such as for instance the predecessor function:
\quad
$ pred = `lx. \cas{\co{0} \mapsto \co{0}; \co{S}
  \mapsto `lz.z}{x}$,\\
which satisfies:\quad
\begin{math}
  \begin{array}[t]{r@{\ = \ }l}
    pred\ (\co{S}\ n) &
    \cas{\co{0} \mapsto \co{0};
      \co{S} \mapsto `lz.z}{(\co{S}\ n})\\
    & (\cas{\co{0} \mapsto \co{0};
      \co{S} \mapsto `lz.z}{\co{S}})\ n\\
    & (`lz.z)\ n \\
    & n
  \end{array}
\end{math}

Actually, one can even encode pattern matching for variadic
constructors.
The $`l$-calculus with constructors enjoys many good properties, such
as confluence and separation (in the spirit of B\"ohm's theorem).
It comprises nine rules, among which we can distinguish
\emph{essential rules} ---such as $`b$-reduction, case analysis and
\rul{CaseApp}--- that are necessary to reduce terms to values, and
\emph{unessential rules} ---like $`h$-reduction--- whose main role is
to guarantee confluence and separation properties.

A polymorphic type system has been proposed for this calculus
in~\cite{Petit09}, thus addressing the problem of typing the case
construct in presence of the \rul{CaseApp} commutation rule.
This paper is an extended version of~\cite{Petit09} with major
changes, since some results appear to be incorrect
(\textit{cf}. Part~\ref{sec:restr}).
Indeed, typed lambda-calculus with constructors supports some
non-terminating reductions, and also match failure can occur.
This is due to one of the unessential rule: the composition between
case constructions.

In this paper we drop out this composition rule from the
calculus\footnotemark, and then justify this with realisability arguments.
\footnotetext{Losing thereby the separation property.}
A semantic analysis using reducibility candidates ensures the strong
normalisation of this restricted calculus.
The main difficulty is to design a good notion of reducibility
candidates which is able to cope with the commutation rule attached to
the case.
For that we introduce the notion of \emph{case commutation normal
  form}, and we consider the usual reducibility
candidates~\cite{girard89} up to case commutation.
From this construction we deduce the main property of the typed
calculus, including the absence of match failure for well typed
terms.

\paragraph*{Outline:}

Parts~\ref{sec:lc} and~\ref{sec:typing} respectively present the
\lc-calculus and the type system.
Part~\ref{sec:restr} is a discussion about the type system and the
different reduction rules,
and Part~\ref{sec:cr} the reducibility candidates model.
Finally, Part~\ref{sec:sn} concludes with the main properties of the
typed \lc-calculus.

\section{The lambda-calculus with constructors}
\label{sec:lc}
\subsection{Its syntax}

The syntax of the $\lambda$-calculus with constructors~\cite{AAAjournal}
is defined from two disjoint sets of symbols: \emph{variables}
(notation: $x$, $y$, $z$, etc.) and \emph{constructors}
(notation:~\co{c}, \co{d},~etc. in typewriter font).
It consists of two syntactic categories defined by mutual induction
in Fig.~\ref{fig:terms}: terms (notation:~$s, t, u$, etc.) and case
bindings (notation:~$`q, `f$).
\begin{figure}[ht]
  \framebox[\textwidth]{
  \begin{math}
    \begin{array}{l@{\quad}lr@{\quad}ll}
      &&&&\\
      Terms:& s,t,u & \triangleq & x \quad|\quad
      `lx.t \quad|\quad tu & (`l\text{-calculus)}\\
      && | & \co{c} & \text{(Constructor)}\\
      && | &\cas{`q}{t} & \text{(Case Construct)}\\
      && | & `? & \text{(Daimon)}\\
      &&&&\\
      Case Bindings: & `q,`f & \triangleq & 
      \{\co{c}_1 \mapsto u_1; \hdots; \co{c}_n \mapsto u_n\} & 
      \text{(Case Binding)}\\
      \multicolumn{5}{c}{\co{c}_i \neq \co{c}_j \text{ for } i \neq j}\\
      &&&&\\
    \end{array}
  \end{math}
}
  \caption{$\lc$-terms and case bindings.}
  \label{fig:terms}
\end{figure}

Terms include all the syntactic constructs of the $`l$-calculus, plus
constructors (as constants) with a case construct (similar to the case
construct of Pascal) to analyse them.
There is also a constant~$`?$ (the \emph{Daimon}, inherited from
ludics~\cite{ludique}) representing immediate termination.
It cannot appear in a term during reduction, but we keep it in the
calculus for technical reasons (explained in
Section~\ref{sec:def-cr}).
Case bindings are finite functions from constructors to terms.
In order to ease the reading, we may 
write~\begin{math}
  \cas{\co{c}_1 \mapsto u_1 \;;\; \hdots\;;\; \co{c}_n \mapsto u_n}{t}
\end{math}
\linebreak[4]
for~\begin{math}
  \cas{\{\co{c}_1 \mapsto u_1; \hdots; \co{c}_n \mapsto u_n\}}{t}
\end{math}.

Free and bound (occurrences of) variables are defined as usual, taking
care that constructors are not variables and thus not subject to
$\alpha$-conversion.
The set of free variables (denoted by $\fv{-}$) is defined for the new
constructs by
\begin{mathpar}
  \inferrule{}{\fv{\co{c}} = \emptyset}\and
  \inferrule{}{\fv{\cas{`q}{t}} = \fv{`q} \cup \fv{t}}\and
  \inferrule{}{\fv{`q} =
    \cup_{(\co{c} \mapsto u)`: `q}\fv{u}} 
\end{mathpar}
A term is \emph{closed} when it has no free variable, and we
write~$\cterms$ for the set of closed \lc-terms.

The usual operation of substitution on terms (notation:
$t\subs{x}{u}$) is defined as expected, taking care of renaming bound
variables when needed in order to prevent variable capture.
Substitution on case bindings (notation: $`q \subs{x}{u}$) is defined
component-wise.

\subsection{Its operational semantics}

The reduction of \lc-calculus is based on the nine reduction rules
given in Fig.~\ref{fig:rules} among which one can find the $`b$ and
$`h$ reduction rules of the $`l$-calculus, now called \rul{AppLam} and
\rul{LamApp} \footnotemark, respectively.
\footnotetext{In \lc-calculus, the name of each reduction rule consists
of the names of the two constructions interacting for the
reduction. }
We write~$\to$ the contextual closure of these rules, and
$\to^=$ (\textit{resp.} $\to^+$, \textit{resp.} $\to^*$) denotes its
reflexive (\textit{resp.} transitive, \textit{resp.} reflexive and
transitive) closure.

Case bindings behave like functions with finite domain.
Therefore we may use the usual functional vocabulary:
if $`q=\{c_i\mapsto u_i\, /\, 1\leq i\leq n\}$, then the \emph{domain}
of~$`q$ is the set $\mathit{dom}(`q)=\{c_1,\dots c_n\}$;
also~$`q_c$ denotes~$u$ \mbox{when $c\mapsto u`:`q$}.
Case constructs are propagated through terms via the \rul{CaseApp,
  CaseLam} and \rul{CaseCase} commutation rules, and ultimately
destructed with \rul{CaseCons} reduction.
For an explanation of the role and expressiveness of these rules,
see~\cite{AAAjournal}.

\begin{figure}[ht]
  \framebox[\textwidth]{
  \begin{math}
    \begin{array}
      {@{\qquad}l@{\quad}l@{\quad\quad}rcl@{\qquad}l}
      &&&&\\
      \multicolumn{6}{l}{\textbf{Beta-reduction}}\\
      \rul{ AppLam} & (\rul{AL}) &
      (`lx.t)u    &"->"& t\subs{x}{u} & \\
      \rul{ AppDai} & (\rul{AD}) &
      `?\,u          &"->"& `? & \\
      &&&&&\\
      \multicolumn{6}{l}{\textbf{Eta-reduction}}\\
      \rul{ LamApp} & (\rul{LA}) &
      `lx.tx &"->"& t & (x\notin \fv{t}) \\
      \rul{ LamDai} & (\rul{LD}) &
      `lx.`? &"->"& `? & \\
      &&&&&\\
      \multicolumn{6}{l}{\textbf{Case propagation}}\\
      \rul{ CaseCons} & (\rul{CO}) &
      \cas{`q}{c} &"->"& t & ((c\mapsto t)\in`q) \\
      \rul{ CaseDai} & (\rul{CD}) &
      \cas{`q}{`?} &"->"& `? & \\
      \rul{ CaseApp} & (\rul{CA}) &
      \cas{`q}{(tu)} &"->"& (\cas{`q}{t})u & \\
      \rul{ CaseLam} & (\rul{CL}) &
      \cas{`q}{`lx.t} &"->"&
      `lx.\cas{`q}{t} & (x\notin \fv{`q}) \\
      &&&&&\\
      \multicolumn{6}{l}{\textbf{Case composition}}\\
      \rul{ CaseCase} & (\rul{CC}) &
      \cas{`q}{ \cas{`f}{t} } &"->"&
      \cas{`q\circ`f}{t} &\\
      \multicolumn{6}{r}{\qquad \text{with }
        `q \circ \{c_1 \mapsto t_1;...; c_n \mapsto t_n\} \equiv
        \{ c_1\mapsto \cas{`q}{t_1};...;c_n \mapsto \cas{`q}{t_n} \}
        \qquad
      }\\
     &&&&&\\
    \end{array}
  \end{math}
}
  \caption{Reduction rules for $\lc$.}
  \label{fig:rules}
\end{figure}

The confluence or non confluence is known for every combination of
the 9 reduction rules (\cite{AAAjournal}~Theorem~1), and the full
calculus is confluent.
In this paper, we shall only consider the following sub-calculi, which
are all confluent:
\begin{enumerate}[$\bullet$]
\item \lcm \, denotes \lc-calculus with all the rules except
  \rul{CaseCase}.
  In this paper we show that types ensure the strong normalisation of
  this calculus.
\item \ccalc \, is the calculus of case commutation (whose only rules
  are \rul{CaseApp} and \rul{CaseLam}).
  For technical reasons (\textit{cf.} Part~\ref{sec:cr}) we sometimes
  consider terms up to case commutation equivalence.
\item \bcalc \, is the complement calculus of \ccalc\ in \lcm:
  it is composed of rules \rul{AppLam}, \rul{AppDai} and
  \rul{LamApp}, \rul{LamDai}, \rul{CaseCons} and \rul{CaseDai}.
\end{enumerate}
A term with no infinite reduction is said to be \emph{strongly
  normalising}.
By extension, a calculus is strongly normalising  when
all its terms are.
It is also known that the whole calculus without \rul{AppLam} is
strongly normalising (\cite{AAAjournal}, Proposition~2).

\subsection{Values in lambda-calculus with constructors}
\label{sec:val}
In pure lambda-calculus, a value is a function (\textit{i.e.} a
$`l$-abstraction).
In \lc\ we call \emph{data structure} a term of the form
$\co{c}\,t_1\hdots t_k$ where~$\co{c}$ is a constructor and
$t_1,\ldots,t_k$ ($k\ge 0$) are arbitrary terms.
We then call a \emph{value} a term which is a $`l$-abstraction or a
data structure.
The set of values is written~\val.

We say that a term is \emph{defined} when it has no sub-term of the
form~$\cas{`q}{\co{c}}$, with $\co{c} `; dom(`q)$, and that it is
\emph{hereditarily defined} when all its reducts (in any number of
steps) are defined.
(Intuitively, non-defined terms contain pattern matching
failures and therefore will be rejected by the type system.)

\begin{prop} \label{prop:fn-val}
  Every defined closed normal term is either $`?$ or a value.
\end{prop}
\proof
  Let $t$ be a closed defined term in normal form.
  By induction on the structure of~$t$, we show that~$t$ is
  either~$`?$ or $`lx.t_0$ or $\co{c}t_1\dots t_k$ for some
  constructor~$\co{c}$, and some terms $t_i$.
  Since~$t$ is closed it is not a variable.
  If it is a constructor, the Daimon or an abstraction, the result
  holds.

  If it is an application, write $h\,t_1 \dots t_k = t$, where~$h$ is not
  an application.
  Then~$h$ is necessarily closed, defined and normal.
  It is not an abstraction, nor the Daimon (otherwise~$t$ would be
  reducible with \rul{AppLam} or \rul{AppDai}).
  Hence it is a data-structure by induction hypothesis, and so is $t$.

  Now assume $t = \cas{`q}{h}$.
  Then~$h$ also is  closed, defined and normal.
  It cannot be the Daimon, nor an abstraction, nor an application,
  otherwise $t$ would be reducible with \rul{CaseDai}, \rul{CaseLam} or
  \rul{CaseApp}.
  So~$h$ is a constructor.
  If it is in the domain of $`q$, then $t$ is reducible with
  \rul{CaseCons}, and if it is not in the domain,~$t$ is not defined.
  Finally~$t$ cannot be a case construct.\qed

Notice that the proof does not use rule \rul{CaseCase} (and rules
\rul{LamApp}, \rul{LamDai} neither), so the proposition holds for
normal forms w.r.t. \lcm.

Finally, a term which is both strongly normalising and hereditarily
defined is said to be \emph{perfectly normalising}.
Perfect normalisation satisfies this usual lemma of lambda-calculus:
\begin{lem}
  \label{lem:pn-subs}
  If $t\subs{x}{u}$ is perfectly normalising, so is~$t$.
\end{lem}
\proof
  First recall that $t \to t'$ implies
  $t\subs{x}{u}\to t'\subs{x}{u}$ (\cite{AAAjournal}~Lemma~9).
  Thus, if $t\subs{x}{u}$ is strongly normalising, so is $t$.
  Then, if $t\subs{x}{u}$ is defined,~it has no sub-term of the form
  $\cas{`q}{c}$ with $c`;\mathit{dom}(`q)$, and this property is 
  kept~by replacing some sub-terms by $x$.
  So $t$ also is defined.
  By induction on the reduction of~$t$, we can easily conclude
  that if~$t\subs{x}{u}$ is hereditarily defined, so is~$t$.\qed

\section{Type system}
\label{sec:typing}
\subsection{An informal presentation}

The type system we want to define includes the simply-typed
$`l$-calculus: the main type construct is the arrow type $T\to U$,
coming with its usual introduction and elimination rules.
To achieve polymorphism, we introduce type variables (written $X$,
$Y$ etc.) and universal type quantification (notation:~$`AX.T$).
Instantiation is performed via a sub-typing judgement containing all
the rules of system $F$ with sub-typing such as presented
in~\cite{mitchell88}.

To type-check data structures, we associate to every
constructor~$\co{c}$ a type constant~$\tco{c}$ ---written with
bold font.
We introduce a \emph{type application} $DT$ for applied structures, so
that we can derive $\co{c}\,\vec{t}:\tco{c}\,\vec{T}$ from
$\vec{t}:\vec{T}$ 
(see~\ref{sec:formal-syst} for more details on vectorial notations).
Nevertheless, the formation of application types has to be restricted.
Indeed, with a typing rule such as
\begin{mathpar}
  \inferrule{t:T \quad u:U}{tu:TU}\and
\end{mathpar}
if~$t$ is a term of type $\mathit{bool} \to U$, and~$u$ a term
of type $\mathit{nat}$, we would be able to type term~$tu$ with
type~$(\mathit{bool}\to U)\,\mathit{nat}$, which may be a nonsense
if~$t$ implements a function expecting only booleans.
Furthermore, it would also enable typing non normalising terms
like~$\delta\delta$, as $`d=`lx.xx$ is typable in system~$F$.

For that reason we distinguish a sub-class of \emph{data types} (notation:
$D$, $E$). They will be the only types on the left-hand side of a type
application.
In practice this sub-class excludes arrow types and type variables
(which could be instantiated by arbitrary types).
To still keep the ability to quantify over data types, we introduce
\emph{data type variables} (notation: $`a$, $`b$ etc.) and data type
quantification.

To encode algebraic types, we add union types.
For example, we could define a type of natural numbers with
the equation
$\nat \equiv \tco{0} \,\cup\, \tco{S} \, (\nat) $
(where \co{0} and \co{S} are constructors)\footnotemark.
\footnotetext{This would require a fixpoint operator, or a double
  sub-typing rule.}
To distribute arrow among union, we also need \emph{intersection types}:
\begin{displaymath}
(\tco{0} \, \cup \, \tco{S}(\nat) ) \to T \equiv 
(\tco{0} \to T) \,\cap\, (\tco{S}(\nat) \to T).
\end{displaymath}
By symmetry, we add the existential quantifier.

\begin{figure}[h]
  \framebox[\textwidth]{
    \begin{math}
      \begin{array}
        {l@{\qquad}r@{\;}r@{\quad}l@{\qquad}r}
        &&&&\\
        Types: & T, U & := & X & \text{(Ordinary type variable)}\\
        && | & `a \quad |\quad \tco{c} \quad |\quad DT 
        & \text{(Data type)}\\
        && | & T "->" U & \text{(Arrow type)}\\
        && | & T `U U & \text{(Union type)}\\
        && | & T \cap U & \text{(Intersection type)}\\
        && | & `A `a.T \quad |\quad `AX. T& \text{(Universal type)}\\
        && | & `E `a.T \quad |\quad `E X.T& \text{(Existential type)}\\
        &&&&\\
        Data\ Types: & D,E & := & `a & \text{(Data type variable)}\\
        && | & \tco{c} \quad|\quad DT & \text{(Data structure)}\\
        && | & D`U E & \text{(Union data type)}\\
        && | & D \cap E & \text{(Intersection data type)}\\
        && | & `A `a. D \quad|\quad `A X. D & \text{(Universal data
          type)}\\
        && | & `E `a. D  \quad|\quad `E X. D & \text{(Existential data
          type)}\\
        &&&&\\
      \end{array}
    \end{math}
  }
  \caption{Types of \lc.}
  \label{fig:types}
\end{figure}

\subsection{The formal system}
\label{sec:formal-syst}

We define a polymorphic type system with union and intersection for
both terms and case bindings of \lc\ (Fig.~\ref{fig:types}).
It uses two spaces of type variables:
\emph{ordinary type variables} 
and \emph{data type variables}. 
There are also two kinds of types:
\emph{ordinary types}, and their syntactic sub-class of \emph{data
  types}. 

In the following, $`n$ denotes a variable which can be an ordinary
type variable or a data type variable.
The set $\tv{T}$ denotes the set of all free type variables of a
type~$T$:
\begin{displaymath}
  \begin{array}[t]{r@{\ =\quad}l@{\qquad}r@{\ =\quad}l}
    \tv{X} & \multicolumn{2}{c}{
     \{X\}\qquad\tv{`a}=\ \{`a\}\qquad\tv{\tco{c}}=~~} & \emptyset\\
    \tv{T\to U} & \tv{T}\cup\tv{U} &
    \tv{DT} & \tv{D}\cup\tv{T} \\
    \tv{T\cap U} & \tv{T}\cup\tv{U} &
    \tv{T\cup U} & \tv{T}\cup\tv{U}\\
    \tv{`A`n.T} & \tv{T}\setminus\{`n\} &
    \tv{`E`n.T} &\tv{T}\setminus\{`n\}
  \end{array}
\end{displaymath}
We also use a vectorial notation for type application and arrow types:
\begin{displaymath}
  \begin{array}[t]{r!{=\;}l!{\quad \quad}r!{\ =\ }l}
    \multicolumn{3}{r}{\vec{T}\ :=\quad `0 \;\; | \;\; \vec{T}; T}&\\
    \multicolumn{4}{r}{}\\
    \;\tco{c}`0 & \tco{c} & \tco{c}(\vec{T};T) & (\tco{c} \vec{T})T\\
    \; `0 \to U & U & (\vec{T}; T) \to U & \vec{T} \to (T \to U)
  \end{array}
\end{displaymath}
Typing rules (Fig.~\ref{fig:typing}) include the usual introduction and
elimination rules of typed $`l$-calculus for each type operator.
Some of them ---like the elimination of universal quantifier---
are indeed sub-typing rules (Fig.~\ref{fig:subtype}).

\begin{figure}[ht]
  \framebox[\linewidth]{
    \begin{tabular}{c}
      \\
      \multicolumn{1}{l}{
        \quad
        \textbf{ \underline {Case Binding:} }
        If $`q=\{\co{c}_i\mapsto u_i \,/\, 1\leq i\leq n\}$ with
        $n \geq 0$.}
      \\ \\
      $\displaystyle \ttrul{Cb}\frac{ \big(`G \these u_{i}: \vec{U_i}
        \to T_i \big)_{i=1}^n }
      {`G \these `q : 
        \tco{c}_{i_0}\vec{U_{i_0}} \to T_{i_0}}
      \scriptstyle (1 ``<= i_0 ``<= n)
      \qquad
      \displaystyle \ttrul{Cb_{\bot}}
      \frac{ \big(`G \these u_{i}: T_i \big)_{i=1}^n }
      {`G \these `q : `A`a.`a \to `AX. X}
      $
      \\
      \rule{\textwidth}{0.1ex}
      \\
      \multicolumn{1}{l}{
        \quad
        \textbf{ \underline {Terms:} }
      }
      \\
      $\displaystyle\ttrul{Init}\frac{-}{`G \these x:T}\;(x:T`:`G)$
      \qquad
      $\displaystyle \ttrul{False} \frac{-}{`G \these `?:T}$
      \qquad
      $\displaystyle \ttrul{Constr}\frac{-}{`G \these \co{c}: \tco{c}}$
      \\ \\
      $\displaystyle   \ttrul{\to intro}\frac{`G , x:T \these t:U}{`G
        \these `lx.t: T"->"U}$
      \qquad \qquad
      $\displaystyle \ttrul{"->"elim}\frac{`G \these t:T"->"U \quad
        `G\these u:T}{`G \these tu: U}$
      \\  \\
      $\displaystyle \ttrul{case}\frac{`G \these t:\vec{U} "->" T
        \qquad `G \these `q: T \to T'}{`G \these \cas{`q}{t}: \vec{U}
        \to T'}$
      \\  
      \rule{\textwidth}{0.1ex}
      \\
      \multicolumn{1}{l}{
      \quad
      \textbf{ \underline {Shared rules:} } $M$ is either a term $t$
      or a case binding $`q$.
    }
    \\  \\
    $\displaystyle \ttrul{Univ} \frac{`G \these M:T}{`G \these M:
      `A`n.T} `n `; \tv{`G}$
    \qquad
    $\displaystyle \ttrul{Inter}\frac{`G \these M:T \quad `G
      \these M:U}{`G \these M: T \cap U}$
    \\  \\
    $\displaystyle \ttrul{Exist} \frac{`G, x:T \these M:U}{`G, x:
      `E`n.T \these M: U} `n `; \tv{U}$
    \quad
    $\displaystyle \ttrul{Union}\frac{`G, x: T_1 \these M:U \quad `G,
      x: T_2 \these M:U}{`G, x: T_1 \cup T_2 \these M: U}$
    \\ \\
    $\displaystyle \ttrul{Subs}\frac{`G \these M:T \quad T`<U}{`G
      \these M:U}$
    \\ \\
  \end{tabular}
}
 \caption{Typing rules}
 \label{fig:typing}
\end{figure}

Type application takes precedence over all the other operators and is left
associative.
Sub-typing rule \srul{Data}\, allows typing constructors with
non-fixed arity:
$$\tco{c}T_1\dots T_k \ `<\ T_{k+1} \to \tco{c}T_1\dots T_kT_{k+1},$$
implies that if $\co{c}t_1\dots t_k$ has type $\tco{c}T_1\dots T_k$,
and if $t_{k+1}$ has type $T_{k+1}$, then $\co{c}t_1\dots t_{k+1}$ has type
$\tco{c}T_1\dots T_{k+1}$.
By iterating, we immediately get
\begin{displaymath}
  (`G\these t_i: T_i)_{i=1}^n
  \qquad\implies\qquad
  `G\these\co{c}t_1\dots t_n:\ \tco{c}T_1\dots T_n
\end{displaymath}
Having such variadic constructors allows for example to add or remove
an element in an array locally (Example~\ref{ex:tableau}).

\begin{figure}[p]
  \framebox[\textwidth]{
    \begin{math}
      \begin{array}{c}
        \\ \\
        \displaystyle
        \strul{Refl} \frac{-}{T `< T}
        \qquad
        \strul{Trans} \frac{T `< T_0 \quad T_0 `< T'}{T `<
          T'}
        \\ \\
        \displaystyle
        \strul{Arrow} \frac{T' `< T \quad U `< U'}{T"->"U
          `<T' "->" U'}
        \qquad\qquad
        \strul{App} \frac{D `< D' \quad T`< T'}{DT `<D'T'}
        \\ \\
        \displaystyle
        \strul{\cup introL} \frac{-}{U_1 `< U_1 `U U_2}
        \qquad
        \strul{\cup introR} \frac{-}{U_2 `< U_1 `U U_2}
        \qquad
        \strul{\cup elim} \frac{T_1 `< U \quad T_2 `<
          U}{T_1 `UT_2 `< U} 
        \\ \\
        \displaystyle
        \strul{\cap intro} \frac{T`< U_1 \quad T`<
          U_2}{T `< U_1
          \cap U_2}
        \qquad
        \strul{\cap elimL} \frac{-}{U_1 \cap U_2 `< U_1}
        \qquad
        \strul{\cap elimR} \frac{-}{U_1 \cap U_2 `< U_2}
        \\ \\
        \displaystyle
        \strul{`A intro} \frac{T`< U}{T `< `A`n.U}
        {\scriptstyle `n `; \tv T}
        \quad
        \strul{`A elim} \frac{-}{`AX.T `< T \{X "<-" U\}}
        \quad
        \strul{`A elimD} \frac{-}{`A`a.T `< T \{`a "<-" D\}}
        \\ \\
        \displaystyle
        \strul{`E intro} \frac{-}{T\{X\! "<-" U\} `< `E X. T}
        \quad
        \strul{`E introD}
        \frac{-}{T\{`a\! "<-" D\} `< `E `a. T}
        \quad
        \strul{`E elim} \frac{U `< T}{`E `n.U `< T}
        {\scriptstyle `n `; \tv T}
        \\ \\
        \displaystyle
        \strul{Data} \frac{-}{D `< T "->" DT}
        \qquad \qquad
        \strul{Constr}
        \frac{-}{\tco{c}_1 \vec{T} \cap \tco{c}_2 \vec{U} `< `A
          `a. `a} {\scriptstyle \co{c}_1 \neq \co{c}_2}
        \\
        \\
        \rule{\textwidth}{0.1ex}
        \\
        \\ 
        \displaystyle
        \strul{App/\cap}  \frac{-}
        {\bigcap_i D_iT_i `< (\bigcap_i D_i) (\bigcap_i T_i)}
        \quad
        \strul{App/`A}   \frac{-}
        {`A `n. (D T) `< (`A `n. D) (`A `n. T)}
        \\ \\
        \displaystyle
        \strul{\to/\cap}  \frac{-}
        {\bigcap_i T_i \to U_i `< (\bigcap_i T_i) \to (\bigcap_i U_i)}
        \quad
        \strul{\to/`A}   \frac{-}
        {`A `n. (T \to U) `< (`A `n.T) \to (`A `n. U)}
        \\ \\
        \displaystyle
        \strul{\to/\cup}  \frac{-}
        {\bigcap_i T_i \to U_i `< (\bigcup_i T_i) \to (\bigcup_i U_i)}
        \quad
        \strul{\to/`E}  \frac{-}
        {`A `n. (T \to U) `< (`E `n.T) \to (`E `n. U)}
        \\ \\
        \displaystyle
        \strul{\cup/AppR} \frac{-}
        {D(\bigcup_i T_i) `< \bigcup_i DT_i}
        \quad
        \strul{\cup/AppL} \frac{-}
        {(\bigcup_i D_i)T `< \bigcup_i (D_iT)}
        \\ \\
        \displaystyle
        \strul{`E/AppR} \frac{-}
        {D(`E`n.T) `< `E`n. DT}  {\scriptstyle `n `; \tv D}
        \quad
        \strul{`E/AppL} \frac{-}
        {(`E`n.D)T `< `E`n. DT} {\scriptstyle `n `; \tv T}
        \\ \\
        \displaystyle
        \strul{\cup/`A}  \frac{-}
        {`A`n. (T \cup U) `< (`A`n. T) \cup U} {\scriptstyle `n `; \tv U}
        \quad
        \strul{`E/\cap}  \frac{-}
        {(`E`n. T) \cap U `< `E`n. (T \cap U)} {\scriptstyle `n `; \tv U}
        \\ \\ \\
      \end{array}
    \end{math}
  }
  \caption{Sub-typing rules.}
  \label{fig:subtype}
\end{figure}  

\subsection{Typing case bindings}
\label{sec:typ-cb}

Types for case bindings are the same as the ones for terms.
A case binding is typed (with rule~\trul{Cb})\ like a function waiting for a
constructor of its domain as argument, up to a possible conversion of
arrow type into application type: \label{page:indec}
from a typing judgement $`G\these u:T\to U$, both following
derivations are valid.
\begin{mathpar}
  \inferrule{`G\these u:T\to U}
  {`G\these\{\co{c}\mapsto u\}:\tco{c}\to T\to U}
  \and
  \inferrule{`G\these u:T\to U}
  {`G\these\{\co{c}\mapsto u\}:\tco{c}T\to U}
\end{mathpar}
This is the point that allows \rul{CaseApp} commutation rule to be
well typed.
\begin{exa}
  \label{ex:tableau}
  Consider the constructor \ctab\ that initialises arrays.
  Then the case binding $`q=\{\ctab\mapsto`lxy.\ctab x\}$ removes
  the second element of any array:
  \begin{displaymath}
    \cas{`q}{(\ctab t_1t_2t_3)}\ \to_{\trul{CA}}^3\ 
    (\cas{`q}{\ctab})t_1t_2t_3\ \to\
    (`lxy.\ctab x)t_1t_2t_3\ \to^3\ \ctab t_1t_3
  \end{displaymath}
  From $\these t_1:T_1$, $\these t_2:T_2$ and $\these t_3:T_3$ we
  can derive $\these \cas{`q}{(\ctab t_1t_2t_3)}: \ttab T_1T_3$:
  \begin{center}
    \AxiomC{}
    \UnaryInfC{$\these `lxy.\ctab x: T_1 \!\to T_2 \!\to \ttab T_1$}
    \AxiomC{}
    \UnaryInfC{$\ttab T_1 `< T_3 \to \ttab T_1T_3$}
    \doubleLine
    \UnaryInfC{$T_1 \!\to T_2 \!\to \ttab T_1 `< 
      T_1 \!\to T_2 \!\to T_3 \!\to \ttab T_1T_3$}
    \BinaryInfC{$\these `lxy.\ctab x: T_1\to T_2\to T_3\to\ttab T_1T_3$}
    \UnaryInfC{$\these `q: \ttab T_1T_2T_3 \to \ttab T_1T_3$}
    \DisplayProof
  \end{center}
  \vspace{0.1cm}
  \begin{center}
    \AxiomC{}
    \UnaryInfC{$\these `q: \ttab T_1T_2T_3 \to \ttab T_1T_3$}
    \AxiomC{$\these t_1:T_1$}
    \noLine
    \UnaryInfC{$\these t_2:T_2$}
    \noLine
    \UnaryInfC{$\these t_3:T_3$}
    \doubleLine
    \UnaryInfC{$\these \ctab t_1t_2t_3: \ttab T_1T_2T_3$}
    \BinaryInfC{$\these \cas{`q}{(\ctab t_1t_2t_3)}: \ttab T_1T_3$}
    \DisplayProof
  \end{center}
We can also give the same type to~$(\cas{`q}{\ctab})t_1t_2t_3$ by
choosing another possible type for~$`q$ (we write~$\vec{T}=T_1;T_2;T_3$):
\begin{center}
    \AxiomC{}
    \UnaryInfC{$\these `lxy.\ctab x:\vec{T}\to\ttab T_1T_3$}
    \UnaryInfC{$\these `q: \ttab\to \vec{T} \to \ttab T_1T_3$}
    \AxiomC{}
    \UnaryInfC{$\these\ctab:\ttab$}
    \BinaryInfC{$\these\cas{`q}{\ctab}:\vec{T} \to \ttab T_1T_3$}
    \AxiomC{\hspace{2cm}$\these t_1:T_1$}
    \noLine
    \UnaryInfC{\hspace{2cm}$\these t_2:T_2$}
    \noLine
    \UnaryInfC{\hspace{2cm}$\these t_3:T_3$}
    \doubleLine
    \BinaryInfC{$\these(\cas{`q}{\ctab})t_1t_2t_3: \ttab T_1T_3$}
    \DisplayProof
  \end{center}
\end{exa}

In the same way, the typing rule (\trul{case}) for a case 
construct~$\cas{`q}{t}$
allows~$t$ to be a function that waits for an arbitrary
numbers of arguments.
This make \rul{CaseLam} well typed.
Indeed, if a case binding~$`q$ has type~$T\to U$, then both terms
$\cas{`q}{`lx.x}$ and $`lx.(\cas{`q}{x})$ are typable with the same
type:
\begin{mathpar}
  \inferrule{x:T\these x:T\and x:T\these`q:T\to U}
  {\inferrule{x:T\these\cas{`q}{x}:U}{\these`lx.(\cas{`q}{x}):T\to U}}
  \and
  \inferrule{\these`lx.x:T\to T \and \these `q:T\to U}
  {\these\cas{`q}{`lx.x}:T\to U}
\end{mathpar}

If the case binding includes many branches, we can either chose one of
them, or give to it an intersection type, and then commute
intersection with arrow.
\begin{exa}
  Assume \nat\ is a type satisfying
  $\nat\equiv\tco{0}\,\cup\,\tco{S}\,\nat$.
  The predecessor case bindings
  $`q=\{\co{0}\mapsto\co{0}\;;\co{S}\mapsto `lx.x\}$ has both types
  $\tco{0}\to\nat$
  and $\tco{S}\,\nat\to\nat$.
  Hence we can derive
  \begin{center}
    \AxiomC{$\these `q:(\tco{0}\!\to\!\nat)\cap(\tco{S}\,\nat\!\to\!\nat)$}
    \AxiomC{$(\tco{0}\!\to\!\nat)\cap(\tco{S}\,\nat\!\to\!\nat)
      `< (\tco{0}\cup\tco{S}\,\nat)\!\to\!\nat$}
    \BinaryInfC{$\these `q:\ (\tco{0}\cup\tco{S}\,\nat)\to\nat$}
    \DisplayProof
  \end{center}
and thus~$`q$ has type~$\nat\to\nat$.
\end{exa}

The rule \trul{Cb_{\bot}} is a kind of generalisation of this typing
derivation:
indeed, if\linebreak[4]
$`q=\{\co{c}_i \mapsto u_i \,/\, 1\leq i\leq n\}$, with
$\these u_i:\vec{U_i}\to T_i$, then for any $J ``(= [1..n]$, the
judgement $\these`q:\bigcup_{i`:J} \co{c}_i\vec{U_i}\to\bigcup_{i`:J}T_i$
is derivable.
Taking $J=\emptyset$, this would be written $\these`q:`A`a.`a\to`AX.X$,
as $`A`a.`a$ is the lower bound of data-types, and $`AX.X$ the lower
bound of types.
In particular,~\trul{Cb_{\bot}} enables typing the empty case binding.
Notice that the only way to type a term~$\cas{\emptyset}{t}$ is
that~$t$ has type~$`A`a.`a$, and this means that~$t$ is (or reduces
on) the Daimon 
(we will see that this is a consequence of
Proposition~\ref{prop:fn-val} and Remark~\ref{rk:adeq-clos}).

\section{Restricted lambda calculus with constructor}
\label{sec:restr}

The type system described in the previous section is the one presented
in~\cite{Petit09}.
It appears that the final result (Proposition~15) of that paper is
wrong\footnotemark.
\footnotetext{In~\cite{Petit09} the proof fails at Lemma~10.
  There is a counterexample to the converse of equivalence (13),
  surprisingly due to the notion of modified substitution used there.}
Here we present a simple counterexample, and we explain how we cope
with the problem.

\subsection{The problem of case-composition}

Typed \lc-calculus does not prevent match failure.
Indeed, the \rul{CaseCase} rule can create sub-terms whose typing is
not checked in the ``dead branches'' of a case-binding.
For instance, if
\\
then \qquad \quad
\begin{math}
  \begin{array}[b]{c@{\qquad \text{and} \qquad}c}
    `f = \{\co{d} \mapsto \co{d'}\} &
    `q = \{\co{c} \mapsto \co{d}\ ; \co{c'} \mapsto \co{c'} \}, \\
    \these `f : \tco{d} \to \tco{d'} & 
    \these `q : \tco{c} \to \tco{d}.
  \end{array}
\end{math}

So we can derive $\these \cas{`q}{\co{c}}: \tco{d}$ and then
$\these \cas{`f}{\cas{`q}{\co{c}}}: \tco{d'}$.
This makes sense because we can obtain
$\cas{`f}{\cas{`q}{\co{c}}}\to^* \co{d'}$ by applying twice the rule
\rul{CaseCons}.
In~$`q$, $\co{c'}\mapsto\co{c'}$ is a dead branch and is forgotten by
the typing (once we know that~$\co{c'}$ itself is typable).
However, we can also apply the rule \rul{CaseCase} and get
$\cas{`f `o `q}{\co{c}}$.
Hence, the second branch of the case-binding is
$\co{c'}\mapsto\cas{`f}{\co{c'}}$, which raises a match failure and is
hardly typable.

The point is that, while typing a case binding, a choice can
implicitly be made concerning the branches that will be taken in
consideration
(if we had chosen type $\tco{c'}\to\tco{c'}$ for $`q$, we would not
have been able to type $\cas{`f}{\cas{`q}{\co{c'}}}$, that reduces on
the same match-failing term $\cas{`f}{\co{c'}}$).
But yet the \rul{CaseCase} rule can create redices in branches that
have been dropped by the typing.

Actually, the situation is even worse.
Rule \rul{CaseCase}, together with the other rules, makes some typable
terms non-terminating:

Let\quad$`f = \{\co{d} \mapsto `d\}$\quad and\quad
$`q = \{\co{c} \mapsto \co{d}\ ; \co{c'} \mapsto \co{d}`d \}$, where
$`d = `lx.xx$.
Then we can derive
\begin{center}
  \AxiomC{$ `G \these `f:  \tco{d} \to `D$}
  \AxiomC{$`G \these \co{d}: \tco{d}$}
  \AxiomC{$`G \these \co{d}`d: \tco{d}`D$}
  \BinaryInfC{$`G \these `q: \tco{c} \to \tco{d}$}
  \AxiomC{$`G \these x: \tco{c}$}
  \BinaryInfC{$`G \these \cas{`q}{x}: \tco{d}$}
  \BinaryInfC{$`G \these \cas{`f}{\cas{`q}{x}}: `D$}
  \DisplayProof
\end{center}
with $`G = x: \tco{c}$, and $`D = (`AX. X \to X) \to (`AX. X \to X)$.
It appears that $\cas{`f}{\cas{`q}{x}}$ is in normal form without
\rul{CaseCase} rule, but with it we can reduce
\begin{displaymath}
  \cas{`f}{\cas{`q}{x}} \to \cas{`f `o `q}{x} =
    \left\{\!\left|
      \begin{array}[c]{l@{\ \mapsto\ }l}
        \co{c} & \cas{`f}{\co{d}}\\
        \co{c'} & \cas{`f}{\co{d} `d}\\
      \end{array}
    \right|\!\right\} \cdot{x}
  \to^*
    \left\{\!\left|
      \begin{array}[c]{l@{\ \mapsto\ }l}
        \co{c} & `d\\
        \co{c'} & `d `d\\
      \end{array}
    \right|\!\right\} \cdot{x}
\end{displaymath}
Hence $\cas{`f}{\cas{`q}{x}}$ is not normalising since the
sub-term~$`d`d$ necessarily appears.

\subsection{Restriction of the calculus}

Remember that \lcm, \textit{i.e.}, the \lc-calculus without the rule~\rul{CaseCase}, is
confluent (\emph{cf}. Part~\ref{sec:lc}).
We will see in Part~\ref{sec:sn} that typed \lcm-calculus enjoys the
perfect normalisation property.

Actually, rule \rul{CaseCase} was introduced in the lambda calculus
with constructors in order to satisfy the separation
property (\cite{AAAjournal},~Theorem~2) ---and same as for the rule
\rul{LamApp}, the usual eta-reduction.
But it is unessential for computing in the lambda calculus with
constructors (cf. the discussion in Section~\ref{sec:discussion}).

Also from now on we remove the case composition from the calculus, and
we consider the \lcm-calculus.
In particular, we now use notation~$\to$ for~$\to_{\lcm}$.

The set of terms is kept unchanged, so we use the same definition of
\emph{defined term} and of \emph{value} as in \lc-calculus.
Note that Proposition~\ref{prop:fn-val} still holds in \lcm.
The set of closed terms that are perfectly normalising for \lcm\ rules
is denoted by~\pn.
By extension we say that a case binding~$`q$ is in \pn\ when it is
composed of closed and perfectly normalising terms for~\lcm.

In the following, we prove the perfect normalisation (\textit{i.e.}
strong normalisation without match failure) of typed
\lcm-calculus.

\section{Reducibility Candidates}
\label{sec:cr}

\emph{Reducibility candidates} \cite{girard89} are sets of closed and perfectly
normalising terms.
They will later be used to interpret types.
In this paper we complete their usual meaning  with the notion of \emph{data
  candidates}.
In the following, we denote by $\red{n}{t}$ the set of terms to which
$t$ reduces in $n$ steps, by $\red{*}{t}$ the union of all these
sets for $n$~in~$\mathbb{N} $, and by $\red{+}{t}$ the union for
$n \geq 1$.

Because of their ``ill-behaviour'' w.r.t. typing, commutation rules
will be treated with a special attention.
Remember that we write~$\cred$ the union of \rul{CaseApp} and \rul{CaseLam},
and \ccalc\ denotes the calculus containing only these two rules.
Conversely, the calculus consisting of all reduction rules of \lcm\
\emph{except} \rul{CaseApp} and \rul{CaseLam} is written \bcalc\ 
(and, as expected, \bred\ denotes the union of \rul{AppLam},
\rul{AppDai}, \rul{LamApp}, \rul{LamDai}, \rul{CaseCons} and \rul{CaseDai}).

In this section, we first give some properties of \ccalc-normal
forms.
Next we give a definition of reducibility candidates and a
method to construct them using closure operator.
Then we emphasise the connection between reducibility candidates and
values.
Finally we define some operations on reducibility candidates.

\subsection{Case-commutation normal form}
The reduction system \ccalc\ is strongly normalising.
Indeed, reducing a term in~\ccalc\ decreases its \emph{structural
  measure}~$s$, introduced in~\cite{AAAjournal} as follows:
\begin{displaymath}
  \begin{array}[t]{c@{\ =\ }l}
    \mes x & \mes c \ =\ \mes{`?} \ =\ 1 \\
    \mes{`lx.t} & \mes t + 1 \\
    \mes{tu} & \mes t + \mes u \\
    \mes{\cas{`q}{t}} & \mes t `* (\mes{`q} + 2)\\
    \mes{\{c_i \mapsto u_i\,/\, 1\leq i\leq n\}} & \sum_{i=1}^n \mes{u_i}
  \end{array}
\end{displaymath}

In the following, we will often need to consider terms up to
case-commutation rules.
The normal form of a term $t$ for \cred\ is written $\sw{t}$.
It is characterised by the following equations:
\begin{displaymath}
  \begin{array}[t]{c@{\ =\ \ }l@{\quad}c@{\ =\ }l}
    \sw x & x & \sw{\cas{`q}{x}} & \cas{\sw{`q}}{x} \\ 
    \sw c & c & \sw{\cas{`q}{c}} & \cas{\sw{`q}}{c} \\ 
    \sw{`?} & `? & \sw{\cas{`q}{`?}} & \cas{\sw{`q}}{`?} \\ 
    \sw{`lx.t} & `lx. \sw t &
    \sw{\cas{`q}{`lx.t}}&`lx.\sw{(\cas{`q}{t})}\\ 
    \sw{(tu)} & \sw t \sw u & \sw{\cas{`q}{(tu)}} &
    \sw{(\cas{`q}{t})} \sw u\\ 
    \sw{\{c_i \mapsto u_i\,/\, {\scriptstyle 1\leq i\leq n}\}} &
    \{c_i \mapsto \sw{u_i}\,/\, {\scriptstyle 1\leq i\leq n}\} &
    \sw{\big(\cas{`q}{\cas{`f}{t}}\big)} &
    \sw{(\cas{`q}{\sw{\cas{`f}{t}}})}
  \end{array}
\end{displaymath}
and by $\sw{\big(\cas{`q}{\cas{`f}{t}}\big)} =\cas{\sw{`q}}{\cas{`f}{t}}$\, 
if\, $\sw{\cas{`f}{t}} \ =\ \cas{`f}{t}$.

To deal with perfect normalisation, we can consider terms up to
case commutation, since both well-definition and strong normalisation
are preserved by \ccalc-reduction and expansion.
That is what Corollary~\ref{cor:pn-cnorm} expresses.
\begin{lem}
  \label{lem:def-cnorm}
  If $\sw t$ is defined, so is $t$.
\end{lem}
\begin{lem}
  \label{lem:red-cnorm}
  $t \bred t'$ implies $\sw t \to^+ \sw t'$
\end{lem}
\proof
  By induction on $t$.
  \begin{enumerate}[$\bullet$]
  \item If $t = x,\ `?$ or~$\co{c}$, then~$t$ is not reducible.
  \item If $t = `lx. t_0$, then $t' = `lx. t_0'$ with
    $t_0 {\bred} t_0'$ and we conclude by induction.
  \item If $t = t_1 t_2 $, three different cases can occur:
    \begin{enumerate}[$-$]
    \item $t' = t_1t_2'$ or $t_1't_2$
      with $t_i {\bred} t_i'$.
      Hence we conclude by induction
    \item $t_1 = `?$ and $t'=`?$.
      In that case $\sw{t} = (`? \sw{t_2})$ reduces to 
      $`? =\ \sw{t'}$.
    \item $t_1 = `lx. t_0$ and $t' = t_0 \subs{x}{t_2}$.
      Then $\sw{t}= (`lx.\sw{t_0}) \sw{t_2}$,
      and it reduces to $(\sw{t_0}) \subs{x}{\sw{t_2}}$, that has
      case normal form (and therefore reduces in~0 or more steps on)
      $\sw{(t_0 \subs{x}{t_2})}$.
    \end{enumerate}
  \item If $t = \cas{`q}{t_0}$, either $t' = \cas{`q'}{t_0}$ or
    $\cas{`q}{t_0'}$ with $`q {\bred} `q'$ or $t_0 {\bred} t_0'$ 
    and we conclude by induction, or $t' = u$ with $t_0 = c$ and
    $c \mapsto u `: `q$, or $t' = `?$ and $t_0 = `?$.
    In both last cases, $\sw{t} = \cas{\sw{`q}}{t_0} \to \sw{t'}$.\qed
  \end{enumerate}
%
\begin{cor}
  \label{cor:pn-cnorm}
  If $\sw t `: \pn$, then $t `: \pn$.
\end{cor}
\proof
  First $u `:\red{*}{t}$ implies $\sw{u} `: \red{*}{\sw{t}}$ by
  Lemma~\ref{lem:red-cnorm}.
  So Lemma~\ref{lem:def-cnorm} entails that all reducts of $t$ are
  defined as soon as all reducts of $\sw{t}$ are.\\
  Now assume there is an infinite reduction
  \begin{math}
    t = t_0 \to t_1 \to t_2 \dots
  \end{math}
    Since \cred\ is strongly normalising, this reduction chain
    contains an infinity of \bred\ reduction steps:
  \begin{math}
    t = t_0 \cred[*] t_{i_1}  \bred\ t_{j_1} \cred[*] 
    t_{i_2}  \bred\ t_{j_2} \dots
  \end{math}
  So $\sw{t_{j_k}} =\ \sw{t_{i_{k+1}}}$ and $\sw{t_{i_k}} \to^+
  \sw{t_{j_k}}$ by Lemma~\ref{lem:red-cnorm}.
  Hence there is an infinite reduction
  \begin{displaymath}
    \sw{t} =\, \sw{t_{i_1}}  \to^+\ \sw{t_{j_1}} =\, \sw{t_{i_2}} 
    \to^+\ \sw{t_{j_2}} =\, \sw{t_{i_3}} \to^+\ \sw{t_{j_3}} \dots
  \end{displaymath}
  This is absurd if \sw{t} is strongly normalising.
  So finally if $\sw{t}$ is perfectly normalising then $t$ also is.\qed

\subsection{Definition of reducibility candidates}
\label{sec:def-cr}

The definition of reducibility candidates is founded on the notion of
values and neutral terms.
Recall that the set \val\ of values includes all data structures and
lambda-abstractions.
We then call \emph{neutral} the terms which are not values.
The set of defined closed neutral terms is written \neu.
In particular,~$`?$ is neutral.
\begin{rem}
  \label{rem:neu-cnorm}
  Since $t `: \val$ implies $\sw t `: \val$, Lemma~\ref{lem:def-cnorm}
  leads to
  $$\sw t `: \neu \implies t `: \neu$$
\end{rem}

A set~$S$ of closed terms is a \emph{reducibility candidate} when it
satisfies:
\begin{enumerate}[\hbox to8 pt{\hfill}]
\item\noindent{\hskip-12 pt\bf\cri:}\ 
   Perfect normalisation: $S ``(= \pn$
\item\noindent{\hskip-12 pt\bf\crd:}\ 
  Stability by reduction: $t `: S "=>"\red{1}{t} ``(= S$
\item\noindent{\hskip-12 pt\bf\crt:}\ 
  Stability by neutral expansion:
  if $t `: \neu$, then $\red{1}{t} ``(= S \;"=>"\; t `: S$
\item\noindent{\hskip-12 pt\bf\crq:}\ 
  Stability by case-commutation:
  if $t \cred t'$, and $t' `: S$ then $t `: S$
\end{enumerate}
We denote by \cre\ the set of all reducibility candidates, and by
\cro\ the conjunction of all four conditions.
The usual stability properties for reducibility candidates are \cri,
\crd\ and \crt.
Property~\crq\ is specific to this type system, and will be necessary in order
to prove the validity of the $\trul{Cb}$ rule.

Note that every reducibility candidate is non empty (it contains $`?$
as neutral term with no reduct).
This will be important when interpreting arrow types.
Moreover \pn\ is in \cre\ (resulting from
Corollary~\ref{cor:pn-cnorm}, \pn\ is stable by \crq).
%

In some of the proofs of this paper we need to use another definition
of reducibility candidates, that is equivalent.
\begin{lem}
  \label{lem:equiv-cr}
  Given $S ``(= \cterms$, we define two new stability properties: 
\begin{enumerate}[\hbox to8 pt{\hfill}]
\item\noindent{\hskip-12 pt\bf\crdb:}\ 
   $t `: S \ "=>"\ \red * t ``(= S$
\item\noindent{\hskip-12 pt\bf\crqb:}\ 
  $\sw t `: S \ "=>"\ t `:  S$
  \end{enumerate}
  Then 
  a reducibility candidate can be characterised by
  $\cri$, $\crdb$, $\crt$ and $\crqb$ since
    \begin{align}
      \label{al:equiv-cr2}
      \crd \ "<=>"\ & \crdb \\
      \label{al:equiv-cr4}
      \crd \land \crq \ "<=>"\ & \crdb \land \crqb
    \end{align}
\end{lem}
\proof\hfill
  \begin{enumerate}
  \item[\eqref{al:equiv-cr2}]
    \crdb\ obviously implies \crd.
    Conversely, if $S$ satisfies \crd\ and $t `: S$, then we can prove
    by induction on $n$ that $t \to^n u$ implies $u `: S$.
  \item[\eqref{al:equiv-cr4}]
    Assume $S$ satisfies \crq.
    If $t$ is a term such that $\sw{t} `: S$, we can see by induction
    on the reduction $t \cred[*]\sw{t}$ that $t `: S$.
    Conversely, if $S$ satisfies \crdb\ and \crqb, then for any
    $t' `: S$ and any $t \cred t'$, we have $\sw{t} =\;\sw{t'}$ is
    in~$S$ by \crdb\ (since $t' \to^* \sw{t'}$), thus $t `: S$ by \crqb.\qed
  \end{enumerate}

\subsection{Closure properties}

A \emph{non-expansed candidate} is a set of terms that satisfies
\cri\ and \crd.
Sets that satisfy \crq\ in addition (or equivalently \crqb) are called
\emph{pre-candidates of reducibility}.
We write \pcr\ for the family of pre-candidates.
For instance $\{\co c\}$ is a pre-candidate for any constructor~$\co c$.
We will see that such a pre-candidates can be closed by \crt\ to obtain
a reducibility candidate.

\begin{defi}
  For $X ``(= \cterms$, we note \clo{X} its closure by \crt. 
  It is defined inductively by
  \begin{displaymath}
    \frac{t `: X}{t `: \clo X} \qquad  \qquad
    \frac{t `: \neu \quad\red 1 t ``(= \clo X}{t `: \clo X}
  \end{displaymath}
\end{defi}
\begin{lem}
  \label{lem:clo-cr}
  If $P\! `:\! \pcr$,\,then $\clo P$\,is\,the\,smallest reducibility
  candidate containing~$P$.
\end{lem}
\proof
  \clo{P} satisfies \crt\ by definition.
  Using the inductive definition, it is immediate to check (by
  induction) that it satisfies \cri\ and \crdb.
  Now we prove by induction that it satisfies \crqb.
  Let $t`: \cterms$ such that $\sw{t} `: \clo{P}$.
  \begin{enumerate}[$\bullet$]
  \item If~$\sw{t}`:P$ then $t `: P$ since~$P`:\pcr$ and thus
    satisfies \crqb.
  \item Else~$\sw{t}`:\neu$ and~$\red{1}{\sw{t}}``(=\clo{P}$.
    In that case,~$t$ also is in~\neu\ (Remark~\ref{rem:neu-cnorm})
    and for all~$u`:\red{1}{t}$, $\sw{u}`:\red{*}{\sw{t}}$
    (by Lemma~\ref{lem:red-cnorm}).
    Moreover,~$\red{*}{\sw{t}}``(=\clo{P}$ by \crdb,
    thus~$\sw{u}`:\clo{P}$.
    By induction hypothesis, it implies that~$u`:\clo{P}$.
    Hence $\red{1}{t}``(= \clo{P}$, so $t `: \clo{P}$ for being
    neutral.
\end{enumerate}
  Finally $\clo{P}$ is a reducibility candidate.
  Moreover, if~$S$ in \cre\ contains $P$, it also contains \clo{P} by~\crt.\qed

In the previous lemma it would not be sufficient to assume that~$P$ is a
non-expansed candidate, to conclude $\clo{P}`:\cre$ (see example below).
We later (in Lemma~\ref{lem:weak-clo}) characterise more precisely
when a non-expansed candidate can be closed to obtain a reducibility
candidate.
\begin{exa}
  Let $t=`ly.\cas{\co c\mapsto\co c}{y}$\ and\ \
  $u=\cas{\co c\mapsto\co c}{`ly.y}$.
  Then $u\cred t$.\\
  The set~$S=\{`lx.t\}$ satisfies \cri\ and \crd\ but $\clo{S}$
  does not satisfy \crq\ since $`lx.u`;\clo{S}$.
  So \clo{S} is not a reducibility candidate.
\end{exa}

Stability under \crt\ also entails that every reducibility candidate
is infinite:
if~$\mathcal{A}$ is a reducibility candidate containing a term~$t$, it
also contains~$\cas{\co c \mapsto t}{\co c}$ 
as a neutral term whose all reducts (by induction on the reduction of~$t$)
are in~$\mathcal{A}$.
So we can construct an infinite increasing 
family of terms of~$\mathcal{A}$.

A \emph{data candidate} is a reducibility candidate whose all values
are data structures.
The sub-class of data candidates, written \dc, will be helpful to
interpret data types.
\begin{rem}
  \label{rem:data-clo}
  Since the closure by \crt\ only adds neutral terms, if $P$
  is a pre-candidate whose all values are data-structures, then
  $\clo{P} `: \dc$.
  In particular $\clo{\{\co c\}}$ is a data candidate for any
  constructor~\co{c}.
\end{rem}

\subsection{Reducibility Candidates and values}

A reducibility candidate is stable under reduction and under expansion
for neutral terms. 
As a consequence, it is entirely determined by its values.
We call \emph{values of a term}~$t$ (or of a set of terms $S$),
and we write \va{t} (\textit{resp.} \va{S}),
the set of values to which~$t$ (\textit{resp.} a term of~$S$) reduces:
\begin{displaymath}
  \va{t} = \red{*}{t}\cap\val
\end{displaymath}

Note that,\val being closed by reduction, \va{S} is a
non-expansed candidate for any set~$S$ of perfectly normalising terms.
However, it is not necessarily a pre-candidate.
Indeed, even if~$\mathcal{A}`:\cre$ it does not insure
$\va{\mathcal{A}}`:\pcr$.
\begin{exa}
  Consider the reducibility candidate~\clo{S}, with
  \begin{displaymath}
    S=\{\ `lx.\cas{\co c \mapsto\co c}{x}\quad ;\quad 
    \cas{\co c\mapsto\co c}{`lx.x}\ \}\ .
  \end{displaymath}
  \va{\clo{S}} is not stable under \crq\ since it does not contain
  $\cas{\co c\mapsto\co c}{`lx.x}$ whereas\linebreak[4]
  \begin{math}
    \cas{\co c\mapsto\co c}{`lx.x}\quad \cred\quad
    `lx.\cas{\co c\mapsto\co c}{x}
  \end{math}
  \ and\ \ $`lx.\cas{\co c\mapsto\co c}{x}`:\va{\clo S}$~.
\end{exa}

Also it is generally not possible to use the closure operator on a set
of values \va{S} to construct a reducibility candidate.
However, the values of a reducibility candidate are, in some extent,
sufficient to define it~(Corollary~\ref{cor:val-cr}).

\begin{lem}
  \label{lem:val-term}
  If $t `: \pn$ and  $\mathcal{A} `: \cre$, then \quad
  \begin{math}
    t `: \mathcal{A} \ "<=>" \ 
    \va{t} ``(= \mathcal{A}
  \end{math} .
\end{lem}
\proof
  The implication is obvious using \crdb.\\
  We prove the converse by induction on the reduction of $t$ (that is
  well-founded for strongly normalising terms).
  Assume $\va{t} ``(= \mathcal{A}$ and prove that $t `: \mathcal{A}$.
  If $t$ is a value it is clear since $t `: \va t$.
  Otherwise $t `: \neu$, and for all $u$ in $\red{1}{t}$, $u `:
  \mathcal{A}$ by induction hypothesis (since $\va{u} ``(= \va{t} ``(=
  \mathcal{A}$).
  So $t `: \mathcal{A}$ by \crt.\qed
%
\begin{cor}
  \label{cor:val-cr}
  Let $\mathcal{A, B}`: \cre$.
  Then $\va{\mathcal{A}} = \va{\mathcal{B}}$ \emph{iff}
  $\mathcal{A = B}$.
\end{cor}
\proof
  We show the implication, the converse is obviously true.
  Let $\mathcal{A, B} ``(= \cre$, such that $\va{\mathcal{A}} =
  \va{\mathcal{B}}$.
  By Lemma~\ref{lem:val-term},
\[
    \begin{array}[t]{c@{\ \mathit{iff}\ }l}
      t `: \mathcal{A}\ & \va{t} ``(= \mathcal{A}
    \\ & \va{t} ``(= \va{\mathcal{A}}\\ & 
    \va{t} ``(= \va{\mathcal{B}} \\ & 
    \va{t} ``(= \mathcal{B} \\ & t `: \mathcal{B}
    \end{array}\eqno{\qEd}
\]

This characterisation of a reducibility candidate by its values will be
used in the next section to prove that our class \cre\ is stable under union.
For that, we also use a sufficient condition described
in~\cite{colin07}: the \emph{principal reduct} property.
\begin{lem}
  \label{lem:p-reduc}
  Every $t `: \neu$ has a reduct (in one step) $u `: \cterms$ such that
  \begin{displaymath}
    t \to^*v\ \land\ v `: \val \qquad "=>" \qquad u \to^*v
  \end{displaymath}
  A term $u$ that satisfies such a property is called a
  \emph{principal reduct} of $t$.
\end{lem}
\proof
  We define inductively, for every $t `: \neu$ \textit{that can reduce
    on a value}, a term $p(t)$:
  \begin{math}
    \begin{array}[t]{c@{\ =\quad }l}
      p( (`lx.t_0) t_1 \dots t_k ) & t_0 \subs{x}{t_1}\ t_2 \dots t_k\\
      p( (\cas{`q}{t_0}) t_1 \dots t_k ) & p(\cas{`q}{t_0}) t_1 \dots t_k\\
      p(\cas{`q}{c}) & u \quad \text{ if } c \mapsto u `: `q\\
      p(\cas{`q}{`lx.t'}) & `lx. \cas{`q}{t'} \\
      p(\cas{`q}{t_1t_2}) & (\cas{`q}{t_1} ) t_2 \\
      p(\cas{`q}{\cas{`f}{t'}}) & \cas{`q}{p(\cas{`f}{t'})}\\
    \end{array}
  \end{math}
  \\
  The point is that when a neutral term reduces on a value, it is
  necessarily by a reduction step performed at the root of the term (a
  so-called \emph{head reduction}).
  The term~$p(t)$ is obtained from $t$ by reducing in head position.
  Every reduction chain leading from~$t$ to a value $v$ begins
  eventually with reductions in sub-terms, and then the head-reduction
  is performed and gives a term $u'$, that reduces on (or is) $v$.
  So to go from $t$ to $u'$ we can first reduce in head position
  and get $p(t)$, and then perform the same reductions in the sub-terms
  to get $u'$.\qed

\subsection{Candidates operators}

Since we aim to interpret types by reducibility candidates,
we need to define all type operations in \cre.
The definition of arrow is standard~\cite{girard89}.
Here we also define the set application:
for $\mathcal{A}, \mathcal{B} ``(= \cterms$,
  \begin{displaymath}
    \begin{array}{c@{\ \triangleq\ }l}
      \mathcal{A \to B} & \{t `: \cterms  \;/\;
      `A u `: \mathcal{A}, \; tu `: \mathcal{B}\}\\
      \mathcal{A B} & \{ tu \;/\; t `: \mathcal{A},\ u`:\mathcal{B}\}
    \end{array}
  \end{displaymath}

It is standard that \cre\ is stable under arrow
(we prove it in Lemma~\ref{lem:op-cr}), as soon as candidates are not
empty (that is the case here, since they all contain~$`?$).
On the other hand, there is no reason for \cre\ to be closed under
application.
Indeed, none of \cri, \crd, \crt\ and \crq\ is preserved by application.
In Lemma~\ref{lem:op-cr}~\eqref{al:app} we see a way to construct a
reducibility candidate by applying candidate to an other one.
The family~\cre\ is naturally closed by intersection.
We use the same method as in~\cite[Corollary~4.12]{colin07} to deduce its
stability under union~\eqref{al:union}.
\begin{lem}
  \label{lem:incl-union}
  For any family
  $(P_i `: \pcr)_{i `: I}$, $\clo{\bigcup P_i} ``(= \bigcup{\clo P_i}$.
\end{lem}
\proof
  By induction on $t `: \clo{\bigcup P_i}$, we show that
  $t `: \clo P_j$ for some $j `: I$.
  \begin{enumerate}[$\bullet$]
  \item If $t `: \bigcup P_i$, then there is $j `: I$ such that
    $t `: P_j ``(= \clo{P_j}$
  \item If $t `: \neu$ and $\red{1}{t} ``(= \clo{\bigcup P_i}$,
    let $u$ be a principal reduct of $t$.
    Then \mbox{$\va{t}=\va{u}$} (Lemma~\ref{lem:p-reduc}).
    Since $u `: \red{1}{t}$, $u `: \clo{P_j}$ for some $j$ by
    induction hypothesis.
    So $\va{u} ``(= \clo{P_j}$ by \crd, and using
    Lemma~\ref{lem:val-term} we get~$t `: \clo{P_j}$.\qed
  \end{enumerate}

\begin{lem}
  \label{lem:weak-clo}
  Let~$S$ be a non-expansed candidate.
  Then $\clo{S}$ is a reducibility candidate if, for any $t,t'`:\cterms$,
  \begin{displaymath}\left.
      \begin{array}[c]{c}
        t \cred t'\\
        t' `: S
    \end{array}
    \right\} \implies t`: \clo{S}
  \end{displaymath}
\end{lem}
\proof
  By definition \clo{S} satisfies \crt.
  The closure operator \clo{\ \cdot\ } preserves \cri\ and \crd, so
  these two properties also hold in \clo{S}.
  Now, we need to prove \crqb.
  Let~$\sw{t} `: \clo{S}$.
  By Corollary~\ref{cor:pn-cnorm},~$\sw{t}`:\pn$ implies~$t`:\pn$.
  We prove by induction on its reduction that $t `: \clo{S}$.
  If $t = \sw{t}$ it is clear;
  else let $t'$ such that $t \cred t' \cred[*] \sw{t}$.
  By induction hypothesis, $t' `: \clo{S}$.
  \begin{enumerate}[$\bullet$]
  \item If~$t'`:S$ then by hypothesis~$t`:\clo{S}$.
  \item Otherwise~$t'`:\neu$ and~$\red{1}{t'}`:\clo{S}$ (by definition
    of the closure operator).
    Hence~$t$ also is in~\neu\ (same as Remark~\ref{rem:neu-cnorm}).
    Moreover, for any $u `: \red{1}{t}$,
    $\sw t \to ^* \sw u$ by Lemma~\ref{lem:red-cnorm}.
    So~$\sw{u}`:\clo{S}$ by \crd, and $u`:\clo{S}$ by induction
    hypothesis.
    Thus~$\red{1}{t}``(=\clo{S}$ and~$t`:\clo{S}$.
  \end{enumerate}
  So~\clo{S} also satisfies \crqb, it is then a reducibility candidate.\qed
%
\begin{lem}
  \label{lem:op-cr}
  Given $(\mathcal{A}_i)$ and $(\mathcal{D}_i)$ families (possibly
  infinite) of \cre\ and \dc\ respectively, 
  $\mathcal{A}`: \cre$, $\mathcal{D}`: \dc$, and $S$ a non-expansed
  candidate that is non-empty,
    \begin{align}
      \label{al:inter}
      \bigcap \mathcal{A}_i`: \cre \qquad &\text{and} \qquad
      \bigcap \mathcal{D}_i `: \dc\\
      \label{al:union}
      \bigcup \mathcal{A}_i`: \cre \qquad &\text{and} \qquad
      \bigcup \mathcal{D}_i `: \dc\\
      \label{al:arrow}
      S \to  \mathcal{A} &\ `:\ \cre\\
      \label{al:app}
      \clo{\mathcal{D} \mathcal{A}}&\ `:\  \dc
    \end{align}
\end{lem}
\proof\hfill
  \begin{enumerate}
  \item[\eqref{al:inter}]
    \cri, \crd, \crt\ and \crq\ are each preserved by intersection, so 
    $\bigcap \mathcal{A}_i$ and $\bigcap \mathcal{D}_i$ are
    reducibility candidates.
    Since values of $\bigcap \mathcal{D}_i$ are values of
    data-candidates, $\bigcap \mathcal{D}_i `: \dc$.
  \item[\eqref{al:union}]
    All candidates~$\mathcal{A}_i$ satisfy \crt, thus
    $\mathcal{A}_i = \clo{\mathcal{A}_i}$ for any $i$.
    So Lemma~\ref{lem:incl-union} says
    that~$\clo{\bigcup\mathcal{A}_i}``(= \bigcup{\mathcal{A}_i}$.
    The converse inclusion also holds by definition,
    so $\bigcup{\mathcal{A}_i} = \clo{\bigcup \mathcal{A}_i}$.
    Moreover, $\bigcup \mathcal{A}_i$ is pre-candidate since \cri,
    \crd\ and \crq\ are preserved by union.
    thus~$\clo{\bigcup\mathcal{A}_i}$ is a reducibility candidate
    (by Lemma~\ref{lem:clo-cr}), and so is $\bigcup \mathcal{A}_i$.\\
    In the same way,~$\clo{\bigcup\mathcal{D}_i}$ is a reducibility
    candidate. 
    By Remark~\ref{rem:data-clo},~$\clo{\bigcup\mathcal{D}_i}`:\dc$.
    \pagebreak[4]
  \item[\eqref{al:arrow}]
    We prove that~$S\to\mathcal{A}$ satisfy all conditions of~\cro:
    \begin{enumerate}[CR1.]
    \item Let $t`: S\to\mathcal{A}$. There exists~$u`:S$,
      and~$tu`:\mathcal{A}``(=\pn$.
      So $t `: \pn$.
    \item Let $t`:S\to\mathcal{A}$ and $t'`:\red{1}{t}$.
      For any $u`:S$, $tu\to t'u$.
      So $tu`:\mathcal{A}$ implies $t'u `: \mathcal{A}$
      since~$\mathcal{A}$ is closed under reduction.
      Hence $t' `: S \to \mathcal{A}$.
    \item For any $t`:\neu$ such that $\red{1}{t}``(=S\to\mathcal{A}$,
      we prove that $u`:S$ implies $tu`:\mathcal{A}$ by induction on
      the reduction of~$u$.
      Since $t `: \neu$, $tu$ is not a data-structure so $tu `: \neu$.
      Furthermore~$t$ is not an abstraction so every reduct of $tu$ is
      either~$`?$ (if $t =`?$), or $t'u$ with $t'`:\red{1}{t}$,
      or~$tu'$ with $u\to u'$.
      In any case it belongs to $\mathcal{A}$:
      $`?$ by \crt, $t'u$ because $t'`:S\to\mathcal{A}$, and~$tu'$ by
      induction hypothesis.
      So $tu`:\mathcal{A}$ by \crt, thus $t`:S\to\mathcal{A}$.
    \item Let~$t\cred t'$ such that~$t'`:S\to\mathcal{A}$.
      For any~$u`:S$, $tu \cred t'u$ and $t'u`:\mathcal{A}$.
      So $tu`:\mathcal{A}$ by \crq\ in~$\mathcal{A}$.
    \end{enumerate}
    Finally $S\to\mathcal{A}$ is a reducibility candidate.
  \item[\eqref{al:app}]
    First notice that $\clo{\mathcal{DA}}=\clo{\mathcal{DA}\cup`?}$
    (since~$`?$ is neutral with no reduct, it is in the closure of any
    set).
    We call~$S$ the set $\mathcal{DA}\cup`?$, and we will first prove
    that it is a non-expansed candidate.
    Then we will prove that~$t'`:S$ and~$t\cred t'$ imply
    $t`:\clo{S}$.
    Also~$\clo{S}`:\cre$ will result from Lemma~\ref{lem:weak-clo}.
    \begin{enumerate}[$-$]
    \item Let $t`:S$.
      If~$t$ is the Daimon, it is perfectly normalising and it has no
      reduct.
      Otherwise, $t=t_1t_2$ with $t_1`:\mathcal{D}$ and
      $t_2`:\mathcal{A}$.
      We show by induction on their reduction that $t`:\pn$ and
      $\red{1}{t} ``(= S$.
      Term~$t_1$ is not an abstraction since it is in a data
      candidate, so every reduct of~$t$ is either~$`?$ (if $t_1=`?$), 
      or a term on the form $t_1't_2$ or $t_1t_2'$ with $t_i \to t_i'$.
      All this reducts are in~$S$, and they are perfectly normalising
      (possibly by induction hypothesis).
      So $\red{1}{t}``(=S$ and~$t`:\pn$.
      Hence~$S$ satisfy \cri\ and \crd.
    \item Let $t\cred t'$ such that $t'`: S$.
      Then $t' = t_1t_2$ with
      $t_1 `: \mathcal{D}$ and $t_2 `: \mathcal{A}$.
      Either $t = t_1't_2$ or $t_1t_2'$ with $t'_i \cred t_i$
      (in that case $t`: \mathcal{DA}$ since $\mathcal{D}$ and
      $\mathcal{A}$ are closed by expansion for \cred),
      or $t = \cas{`q}{(t_0 t_2)}$ and $t_1 = \cas{`q}{t_0}$.
      In the last case, $t`:\neu$:
      both $\cas{`q}{t_0}$ and~$t_2$ are defined (they are in
      reducibility candidates) so~$\cas{`q}{(t_0t_2)}$ also is
      defined, and it is not a value.
      We show that all its reducts are in \clo{S}. 
      Note that $t_0$ is not an abstraction (if $t_0 = `lx. t_0'$ then
      $t_1\to`lx.\cas{`q}{t_0'}`;\mathcal{D}$), so a reduct~$u$ of~$t$
      may have three different forms:
      \begin{enumerate}[$\bullet$]
      \item $u=t'$.
        Hence $u`:S``(=\clo{S}$.
      \item $u = \cas{`q}{`?}$ (if $t_0 = `?$).
        In that case $u`:\neu$ and all its reducts in any
        number of steps until $`? $ are in \neu, so $u$ is in \clo{S}.
      \item $u=\cas{`q'}{(t_0't_2')}$ with $`q\to`q'$ and $t_i=t_i'$,
        or $`q=`q'$ and $t_i\to t_i'$.\\
        In that case, $u \cred u' = (\cas{`q'}{t_0'})t_2'$, and
        $t'\to u'$ so $u'`:S$ by \crd.
        Thus $u `: \clo{S}$ by induction hypothesis.
      \end{enumerate}
      Hence any reduct of~$t$ is in~$\clo S$, and thus $t`:\clo S$ by \crt.
    \end{enumerate}
    By Lemma~\ref{lem:weak-clo}, $\clo{\mathcal{DA}} = \clo{S} `: \cre$.
    What is more, all values of $\clo{\mathcal{DA}}$ are in
    $\mathcal{DA}$, thus they are applications, so they are
    data-structures.
    Finally, $\clo{\mathcal{DA}} `: \dc$.\qed
  \end{enumerate}

In~\eqref{al:app} we consider the closure of set application for a
data-candidate and a candidate.
In general, the closure of the application of two reducibility
candidates would \emph{not} form a reducibility candidate, as shown in
the following example.
This is intuitively due to the same reason why we do not consider
general type application, but we restrict it to data-types:
good properties (among which the perfect normalisation property) are
insured to be preserved by applying a term~$u$ to~$t$ if~$t$ is not
(and does not reduce on) an abstraction.

\begin{exa}
  Consider the reducibility candidate
  \begin{math}
    \mathcal{A}= \clo{\{I\}},
  \end{math}
  where $I=`lx.x$\\
  Then $II `:\clo{\mathcal{AA}}$, but $II \to I$ and
  $I`;\clo{\mathcal{AA}}$.
  Thus $\clo{\mathcal{AA}}$ is not closed under \crd~and thereby is
  not a reducibility candidate.
\end{exa}

\section{Reducibility model}
\label{sec:sn}
In this section we associate to every type~$T$ a
reducibility candidate that contains all the terms which are typable
by~$T$.
Seeing typed terms as terms of a reducibility candidate or a
data-candidate will then enable a finer analysis of their properties.

\subsection{Modelling types}

To achieve the definition of type interpretation, we need to give the
interpretation for type variables.
For that, we use \textit{valuations}, i.e. functions matching every
data-type variable to a data-candidate, and every type variable to a
reducibility candidate.

Given a valuation $`r$, the \emph{interpretation} of a type
$T$ in $`r$, written $\itp{T}{`r}$, is defined inductively in
Fig.~\ref{fig:interpretation}.
We also associate to $T$ (seen as a type for case bindings) and $`r$
the set of case bindings $\itpcb{T}{`r}$.
Lemma~\ref{lem:op-cr} ensures that for every
valuation~$`r$, $\itp{T}{`r} `: \cre$ for any type $T$, and
$\itp{D}{`r} `: \dc$ for any data type~$D$.

\begin{figure}[ht]
  \begin{math}
    \framebox[\linewidth]{
      \begin{array}[t]{@{\vspace{0.3em}}r!{=}l!{\qquad}r!{=}l}
        \multicolumn{4}{c}{}\\
        \multicolumn{4}{l}{\textbf{Type interpretation by
            reducibility candidates:}}\\
        \itp{`a}{`r} & `r (`a)
        &
        \itp{T \cap U}{`r} & \itp{T}{`r} \cap \itp{U}{`r}
        \\
        \itp{X}{`r} & `r (X)
        &
        \itp{`A`a. U}{`r} & \bigcap_{\mathcal{A} `:\dc}\;\itp{U}{`r,
          `a \mapsto \mathcal{A}}
        \\
        \itp{c}{`r} & \clo{\{c\}}
        &
        \itp{`AX. U}{`r} & \bigcap_{\mathcal{A} `:\cre}\;\itp{U}{`r,
          X \mapsto \mathcal{A}}
        \\
        \itp{DT}{`r} & \clo{ \itp{D}{`r} \ \itp{T}{`r}}
        &
        \itp{T \cup U}{`r} & \itp{T}{`r} \cup \itp{U}{`r}
        \\
        \itp{T \to U}{`r} & \itp{T}{`r} \to \itp{U}{`r}
        &
        \itp{`E`a. U}{`r} & \bigcup_{\mathcal{A}`:\dc}\;\itp{U}{`r, `a
          \mapsto \mathcal{A}}
        \\
        \multicolumn{2}{c}{}
        &
        \itp{`EX. U}{`r} & \bigcup_{\mathcal{A}`:\cre}\;\itp{U}{`r, X
          \mapsto \mathcal{A}}
        \\
        \multicolumn{4}{c}{}\\
        \multicolumn{4}{l}{\textbf{Interpretation of types for case
            bindings:}}\\
        \itpcb{T}{`r}&
        \multicolumn{3}{l}{
          \{ \, `q \, / \, `lx.\, \cas{`q}{x} `: \itp{T}{`r}\}
        }\\
        \multicolumn{4}{c}{}\\
      \end{array}
    }
  \end{math}
  \caption{Interpretation of types}
  \label{fig:interpretation}
\end{figure}
Note that we need to use the closure operator to interpret data
types.
Indeed, for $\mathcal{D}`:\dc$ and $\mathcal{T}`:\cre$, the
set~$\mathcal{DT}$ does not satisfy~\crt:
if~$t`:\mathcal{D}$\ \ and~$u`:\mathcal{T}$, with both terms in normal
form, then the only reduct (assuming~$t\neq`?$) of the
term~$\cas{c\mapsto tu}{c}$\ \ is~$tu`:\mathcal{DT}$,
but~$\cas{c\mapsto tu}{c}$ itself is not an application, and thus is
not in~$\mathcal{DT}$.
However, this interpretation of types gives a very precise notion of
data-types, considering their values.
\begin{prop}
  \label{prop:val-dc}
  If $t$ is a value of $\itp{\tco{c} T_1 \dots T_k}{`r}$
  then $t = \co{c} t_1 \dots t_k$ with \mbox{$t_i`:\itp{T_i}{`r}$}.
\end{prop}
In particular, Proposition~\ref{prop:fn-val} ensures that
$t `: \itp{\tco{c} T_1\hdots T_k}{`r}$ implies 
$t \to^* \co{c}t_1 \hdots t_n$ for some $t_i `: \itp{T_i}{`r}$
($i \leq n$), or $t \to^* `?$.

\proof
  We proceed by induction on $k$.
  
  If $k= 0$, it is straightforward from the definition
  of~$\itp{\tco{c}}{`r}$.
  \\
  Else
  \begin{math}
    \itp{\tco{c} T_1 \dots T_k}{`r} = 
    \clo{ \itp{\tco{c} T_1 \dots T_{k-1}}{`r} \itp{T_k}{`r}}
  \end{math},
  so
  \begin{displaymath}
    \va{\itp{\tco{c} T_1 \dots T_k}{`r}} =
    \va{\itp{\tco{c} T_1 \dots T_{k-1}}{`r} \itp{T_k}{`r}}
  \end{displaymath}
  So, if~$t$ is a value of $\itp{\tco{c} T_1 \dots T_k}{`r}$ it is on the
  form $u u'$ with $u `: \itp{\tco{c} T_1 \dots T_{k-1}}{`r}$ and 
  $u' `: \itp{T_{k}}{`r}$.
  Moreover, if~$uu'$ is a value, it is necessarily a data structure,
  and~$u$ also is a data structure.
  Hence $u$ is a value of $\itp{\tco{c} T_1 \dots T_{k-1}}{`r}$.
  By induction hypothesis $u = \co{c} t_1 \dots t_{k-1}$ with
  $t_i `: \itp{T_i}{`r}$, and we conclude with
  $t_k=u' `: \itp{T_k}{`r}$.\qed
%
\begin{cor}
  \label{cor:data-itp}
  For any constructor $\co{c}$ and any types $T_1, \dots, T_k$,
  \begin{displaymath}
    \itp{\tco{c} T_1 \dots T_k}{`r} \ =\
     \clo{\tco{c} \itp{T_1}{`r} \dots \itp{T_k}{`r}}
  \end{displaymath}
\end{cor}
\proof
    By Proposition~\ref{prop:val-dc},
    $\va{\itp{\tco{c} T_1 \dots T_k}{`r}} =
    \tco{c} \itp{T_1}{`r} \dots \itp{T_k}{`r}$.\\
    Since $\va{\clo{c \itp{T_1}{`r} \dots \itp{T_k}{`r}}}$ also is
    $\tco{c} \itp{T_1}{`r} \dots \itp{T_k}{`r}$, Corollary~\ref{cor:val-cr}
    entails the equality.\qed

The following lemma expresses that type interpretation is sound
w.r.t. sub-typing.
\begin{lem}
  \label{lem:sub}
  If $\ T_1 `< T_2\ $ then for any valuation $`r$,
  $\ \itp{T_1}{`r} \subseteq \itp{T_2}{`r}\ $.
\end{lem}
\proof
  By induction on the derivation of $T_1 `< T_2$.
  Rules \srul{Refl} and \srul{Trans} are straightforward from the
  definition.
  So are union and intersection rules.
  Introduction and elimination rules for quantifiers~$`A$ and~$`E$ use
  the equality
  $\itp{T}{`r, `n \mapsto \itp{U}{`r}} = \itp{T \{`n "<-"U\}}{`r}$.
  \\
  \srul{Arrow} is standard, and \srul{Constr} comes from
  Proposition~\ref{prop:val-dc}:
  $\itp{\tco{c}_1\vec{T}}\cap\itp{\tco{c}_2\vec{U}}{`r}$
  has no value if $\co{c}_1 \neq \co{c}_2$ and thus is smallest than any
  candidate.\\
  We detail rules \srul{App} and \srul{Data},
  other rules are easy to check (we actually introduced them in the
  calculus because they were valid in the  model).\medskip

\begin{enumerate}[\hbox to8 pt{\hfill}]
\item\noindent{\hskip-12 pt\srul{App}:}\
    \begin{math}
      \displaystyle
      \frac{D `< D' \quad T`< T'}{DT `<D'T'}
    \end{math}\bigskip

    \noindent Remark that $ \mathcal{D} ``(= \mathcal{D'}$
    and $ \mathcal{T} ``(= \mathcal{T'}$ implies
    $\mathcal{DT} ``(= \mathcal{D'T'}$,
    and notice that the closure operator is monotone on sets of
    terms.\bigskip

\item\noindent{\hskip-12 pt\srul{Data}:}\
    \begin{math}
      D `< T "->" DT
    \end{math}\medskip

    \noindent Let $`r$ a valuation and $t `: \itp{D}{`r}$.
    Now choose $u`:\itp{T}{`r}$.
    Then $tu `: \itp{D}{`r} \itp{T}{`r}$, and this set
    is included in $ \clo{\itp{D}{`r} \itp{T}{`r}} = \itp{DT}{`r}$.
    Hence~$tu`:\itp{DT}{`r}$ for all~$u$ in~$\itp{T}{`r}$,
    so~$t`:\itp{T\to DT}{`r}$.\qed
  \end{enumerate}

\subsection{Adequacy lemma.}

In this part we prove adequacy for the model: 
if a \lc-term has type~$T$, then it belongs to the
interpretation of~$T$ (and thus is perfectly normalising).

Reducibility candidates model deals with closed terms, whereas
proving the adequacy lemma by induction requires the use of open terms
--- with some assumptions on their free variables, that will be
guaranteed by a context.
Therefore we use \emph{substitutions} $`s$, $`t$ to close terms and
case bindings:
  \begin{displaymath}
    `s := \emptyset \ |\ x \mapsto u;`s \qquad \qquad
    M_{\emptyset} = M; \quad M_{x \mapsto u;`s} = M\subs{x}{u}_{`s},
  \end{displaymath}
We complete the interpretation of types with the one of judgements:
given a context $`G$, we say that a substitution $`s$
\emph{satisfies}~$`G$ for the valuation $`r$ (notation
$\acc{`s}{`G}{`r}$) when $(x:T) `: `G$ implies
$`s(x) `: \itp{T}{`r}$.
A typing judgement $`G \these t: T$ (or $`G \these `q: T$) is said to
be \emph{valid} (notation: $`G \valid t:T$ or $`G \valid `q:T$
respectively) if for every valuation $`r$ and every substitution
$\acc{`s}{`G}{`r}$,
  \begin{displaymath}
    t_{`s} `: \itp{T}{`r}
    \qquad \qquad
    (resp. \, `q_{`s} `: \itpcb{T}{`r})
  \end{displaymath}
The proof of adequacy requires a kind of inversion lemma for \cre.
Recall that $\red{*}{t}$ denotes the set of all reducts (in any number
of steps) of a term~$t$.
\begin{lem}
  \label{lem:inv-cr}
  For any $\mathcal{A} `: \cre$, any terms $t,u,`lx.t_0$, and every
  non-empty non-expansed candidate $S$,
  \begin{align}
    \label{al:inv-app}
    tu `: \mathcal{A}  
    \quad &"<=>" \quad 
    t `: \red{*}{u} \to \mathcal{A}\\
    \label{al:inv-abs}
    `lx.t_0 `: S \to \mathcal{A}  
    \quad &"<=>" \quad 
    \text{for all } s `: S, \, t_0 \subs{x}{s} `: \mathcal{A}
  \end{align}
\end{lem}
\proof\hfill
  \begin{enumerate}
  \item[\eqref{al:inv-app}]
    If $tu `: \mathcal{A}$ then for any $u' `: \red{*}{u}$,
    $tu \to^* tu'$ hence $tu' `: \mathcal{A}$ by \crdb.
    So $t `: \red{*}{u} \to \mathcal{A}$.
    Conversely, if $t `: \red{*}{u} \to \mathcal{A}$ then
    $tu `: \mathcal{A}$ since $u `: \red{*}{u}$.
  \item[\eqref{al:inv-abs}]
    If $`lx.t_0 `: S \to \mathcal{A}$, then for any $s `: S$,
    $(`lx.t_0)s `: \mathcal{A}$, so $(`lx.t_0)s \to  t_0\subs{x}{s}$ implies
    $t_0\subs{x}{s} `: \mathcal{A}$ by \crd.
    Now, if $t_0\subs{x}{s} `: \mathcal{A}$ for some $s `: S$, then
    $t_0`:\pn$ by Lemma~\ref{lem:pn-subs}.
    Moreover, For any $s'`:S$, we can easily check by induction on the
    reduction of~$t_0$ and~$s'$ that $(`lx.t_0)s'`: \mathcal{A}$;
    indeed, it is in \neu, and all its reducts are in~$\mathcal{A}$.\qed
  \end{enumerate}

\begin{rem}
  \label{rk:inv-app}
  If $u `: \pn$, then $\red{*}{u}$ is a non-expansed candidate,
  and so \linebreak[4]
  $\red{*}{u}\to\mathcal{A}`:\cre$ by~\eqref{al:arrow}.
  Also, if $u_i `: \pn$ for $1 \leq i \leq k$, then
\begin{displaymath}
    t\,u_1 \dots u_k `: \mathcal{A}  
    \quad "<=>" \quad 
    t `: \red{*}{u_1} \to \dots \to \red{*}{u_k} \to \mathcal{A}
\end{displaymath}
directly results from~\eqref{al:inv-app} and an
induction on~$k$.
\end{rem}

\begin{lem}
  \label{lem:adeq-case}
  Let $\mathcal{A}_1,\dots,\mathcal{A}_k,\mathcal{B}`:\cre$ 
  and $`q`:\pn$ 
  Assume $\co{c} \mapsto u `: `q$, with
  $u`:\vec{\mathcal{A}}\to\mathcal{B}$
  (where $\vec{\mathcal{A}}=\mathcal{A}_1;\dots;\mathcal{A}_k$).
  Then 
  \begin{displaymath}
    t`:\clo{\tco{c} \mathcal{A}_1\dots\mathcal{A}_k}
    \quad\implies\quad
    \cas{`q}{t} `: \mathcal{B}
  \end{displaymath}
\end{lem}
\proof
  We prove that for all~$`q`:\pn$ with $\co{c}\mapsto u`:`q$ and
  $u`:\vec{\mathcal{A}}\to\mathcal{B}$, and for all
  $t`:\clo{\tco{c}\mathcal{A}_1\dots\mathcal{A}_k}$, the
  term~$\cas{`q}{t}$ is in~$\mathcal{B}$.
  
  If $t$ is a value then $t = \co{c} t_1 \dots t_k$ with
  $t_i`:\mathcal{A}_i$, so
  \\
  \begin{math}
    \begin{array}[t]{l@{\ \text{iff}\quad}l@{\qquad}r}
      \cas{`q}{t} `: \mathcal{B} &
      (\cas{`q}{\co{c}})t_1\dots t_k`:\mathcal{B} & \crdb, \crqb \\
      & \cas{`q}{\co{c}} `: \red{*}{t_1} \to \dots \to \red{*}{t_k}
      \to \mathcal{B} &\text{(Remark~\ref{rk:inv-app})}\\
    \end{array}
  \end{math}
  \\
  But $\red{*}{t_i} ``(= \mathcal{A}_i$, so
  $\mathcal{A}_1 \to \dots \to \mathcal{A}_k \to \mathcal{B}
  ``(= \red{*}{t_1} \to \dots \to \red{*}{t_k} \to \mathcal{B}$.
  Moreover an immediate induction on the reduction of~$`q$ ensures
  that
  $\cas{`q}{\co{c}}$ is in~$\vec{\mathcal{A}}\to\mathcal{B}$:
  this term is in \neu\ and its reducts are either
  $\cas{`q'}{\co{c}}$ \mbox{with $`q \to`q'$}
  (that is in $\vec{\mathcal{A}} \to \mathcal{B}$ by induction
  hypothesis), or~$u$ (that is
  in~$\vec{\mathcal{A}}\to\mathcal{B}$ by hypothesis).
  So~$\cas{`q}{\co{c}}$ is in~$\vec{\mathcal{A}}\to\mathcal{B}$ by~\crt,
  thus it belongs to~$\red{*}{t_1}\to\dots\to\red{*}{t_k}\to\mathcal{B}$
  and so $\cas{`q}{t} `: \mathcal{B}$.
  
  Now assume $t$ is neutral.
  It has the form $h t_1 \dots t_n$ with $h=`?$
  or $\cas{`f}{h_0}$ and $n \geq 0$, or $h = `lx.h_0$ and $n \geq 1$.
  We prove that $\cas{`q}{t}$ is in~$\mathcal{B}$ by induction on the
  reductions of~$`q$ and~$h$.
  \begin{enumerate}[$\bullet$]
  \item 
    First consider cases $h = `?$ or $\cas{`f}{h_0}$, and~$n\geq 0$:
    \\
    \begin{math}
      \begin{array}[t]{l@{\ \text{iff}\quad}l@{\quad}r}
        \cas{`q}{t} `: \mathcal{B} &
        (\cas{`q}{h})t_1\dots t_k`:\mathcal{B} & \crdb, \crqb \\
        & \cas{`q}{h} `: \red{*}{t_1}\!\to\cdot\!\cdot\!\cdot\to\red{*}{t_k}
        \to \mathcal{B} &\eqref{al:inv-app}\\
      \end{array}
    \end{math}
    \\
    Note that $\cas{`q}{h} `: \neu$ and
    $\red{*}{t_1} \to \dots \to \red{*}{t_k} \to \mathcal{B}$ is a
    reducibility candidate by~\eqref{al:arrow}.
    So it is sufficient to show that it contains all reducts of
    $\cas{`q}{h}$.
    They are either $`?$, or $\cas{`q'}{h'}$ with $`q \to `q'$ and
    $h = h'$ or $h \to h'$ and $`q = `q'$.
    The Daimon is in every reducibility candidate, and
    $\cas{`q'}{h'}`:\red{*}{t_1}\to\dots\to\red{*}{t_k}\to\mathcal{B}$
    by induction hypothesis.
    So
    $\cas{`q}{h}`:\red{*}{t_1}\to\dots\to\red{*}{t_k}\to\mathcal{B}$
    by \crt, and $\cas{`q}{t} `: \mathcal{B}$.
  \item
    Now consider case $h=`lx.h_0$ (with $x`;\fv{`q}$), and $n\geq 1$.
    \\
    \begin{math}
      \begin{array}[t]{l@{\ \text{iff}\ }l@{}r}
        \cas{`q}{t}\!`:\mathcal{B} &
        (`lx.\cas{`q}{h_0})t_1\dots t_k`:\mathcal{B} & \crdb,\crqb \\
        & `lx.\cas{`q}{h_0}`: \red{*}{t_1} \!\to\! \cdots \!\to\! \red{*}{t_k}
        \!\to\! \mathcal{B} &\eqref{al:inv-app}\\
        & f\!or\ all\ s `: \red{*}{t_1}, & \\
        \multicolumn{1}{c}{}&\!
        \cas{`q}{\!h_0}\subs{x}{s}\!`:\!
        \red{*}{t_2\!}\!\to\!\cdot\!\cdot\!\cdot\!\to\!\red{*}{t_k\!}\!
        \to\!\mathcal{B} &
        \eqref{al:inv-abs}
      \end{array}
    \end{math}
    \\
    For any $s `: \red{*}{t_1}$,
    $t\to^*(`lx.h_0)\,s\,t_2\dots t_n\to h_0\subs{x}{s}\,t_2\dots t_n$,
    so that \linebreak[4]
    $\cas{`q}{(h_0\subs{x}{s}t_2\dots t_n)} `: \mathcal{B}$ by
    induction hypothesis.
    \\
    Hence, $(\cas{`q}{h_0\subs{x}{v}})t_2\dots t_n `: \mathcal{B}$
    by~\crdb, and thus by~\eqref{al:inv-app},
    $\cas{`q}{h_0} \subs{x}{v}$ belongs to
    $\red{*}{t_2}\to\dots\to\red{*}{t_k}\to\mathcal{B}$.
    Also~$\cas{`q}{t}`:\mathcal{B}$.
  \end{enumerate}
  Finally, $\cas{`q}{t}$ always belongs to $\mathcal{B}$.\qed

\begin{prop}
  \label{prop:adequacy}
  Given a term $t$, a case binding $`q$, a context $`G$ and a
  type $T$,
  \begin{align}
    `G \these t: T \qquad &"=>" \qquad `G \valid t: T\\
    `G \these `q: T \qquad &"=>" \qquad `G \valid `q: T
  \end{align}
\end{prop}
\proof
  The proof is made by induction on the derivation of $`G \these
  t: T$ or $`G \these `q: T$.
  If the judgement is introduced by the rule \trul{Init, False}
  (remember that $`?$ is in every reducibility candidate) or
  \trul{Constr} it is obvious.
  If it comes from \trul{\to elim} it is a direct consequence of the
  definition of arrow in \cre, and the case \trul{\to intro} is a
  consequence of~\eqref{al:inv-abs}.
  \\
  If it comes from \trul{Inter}, \trul{Union}, or \trul{Univ} it is
  straightforward from induction hypothesis.
  If it comes from \trul{Subs}, it is a consequence of Lemma~\ref{lem:sub}.
  We detail the proof in case the derivation comes from rule
  \trul{CB} or \trul{Exist} (\trul{Inter} is similar to this
  last one).
\begin{enumerate}[\hbox to8 pt{\hfill}]
\item\noindent{\hskip-12 pt\bf Cb:}\ 
    \begin{math}\displaystyle
      \frac{(`G \these u_j : \vec U_j \to T_j)_{j=1}^n}
      {`G \these `q: \tco{c}_i \vec{U_i} \to T_i}
    \end{math}
    \quad with $`q = \{\co{c}_j \mapsto u_j \,/\,1\leq j\leq n\}$
    
    \vspace{8pt}
    \noindent
    Remember that the interpretation of a type~$T$, seen as a type for
    case bindings is\linebreak[4]
    \begin{math}
      \itpcb{T}{} = \{`q\ /\ `lx.\cas{`q}{x}`:\itp{T}{}\}
    \end{math}.
    Note $(U_{i1}\dots U_{ik}) = \vec{U_i}$,
    choose~$`r$ a valuation and $\acc{`s}{`G}{`r}$, and show that
    $`lx. \cas{`q_{`s}}{x} `: \itp{\tco{c}_i \vec U_i \to T_i}{`r}$.
    Let $t `: \itp{\tco{c}_i \vec U_i}{`r}$.
    By induction on the reduction of $`q_{`s}$ and $t$, we show that
    $(`lx. \cas{`q_{`s}}{x})t `: \itp{T_i}{`r}$.
    This is a neutral term, so it is sufficient to show that all its
    reducts are in $\itp{T_i}{`r}$.
    Thanks to induction hypothesis we just have to consider the reduct
    $\cas{`q_{`s}}{t}$.\\
    By Corollary~\ref{cor:data-itp}, 
    $t `: \clo{\tco{c}_i \itp{U_{i1}}{`r} \dots \itp{U_{ik}}{`r}}$,
    and $u_i `: \itp{U_{i1}}{`r}\to \dots \to \itp{U_{ik}}{`r} \to
    \itp{T_i}{`r}$ by induction hypothesis.
    All terms in $`q_{`s}$ are perfectly normalising, so we can use
    Lemma~\ref{lem:adeq-case} to get $\cas{`q_{`s}}{t} `:
    \itp{T_i}{`r}$.\bigskip

\item\noindent{\hskip-12 pt\bf Exist:}\ 
    \begin{math}
      \displaystyle
      \frac{`G, x:T \these t:U}{`G, x: `E`n.T \these t: U}
      \scriptstyle `n `; \tv{U}
    \end{math}
    
    \vspace{8pt}
    \noindent Choose a valuation $`r$, and a substitution 
    $\acc{`s}{`G,  x: `E`n.T}{`r}$.\\
    Then $`s(x)`:\bigcup_{\mathcal{A}`:\cre}
    \itp{T}{`r,`n\mapsto\mathcal{A}}$.
    Let $\mathcal{A} `: \cre$.
    Then $`s(x) `: \itp{T}{`r, `n \mapsto \mathcal{A}}$,\linebreak
    so $\acc{`s}{`G, x:T}{`r, `n \mapsto \mathcal{A}}$.
    By induction hypothesis, $(`G, x:T) \valid t:U $,
    so $t_{`s}`: \itp{U}{`r, `n \mapsto \mathcal{A}}$.
    Since $`n `; \tv{U}$, it means that $t_{`s}`: \itp{U}{`r}$.\qed
  \end{enumerate}

\begin{rem}
  \label{rk:adeq-clos}
  For a closed term~$t$ and a closed type~$T$ we immediately get
  \begin{align}
    \label{al:adequacy}
    \itp{T}{}`:\cre \qquad \text{ and } \qquad
    \these t:T \ "=>"\ t`:\itp{T}{}
  \end{align}
\end{rem}

\subsection{Results from the model}
\label{sec:discussion}

Remembering that reducibility candidates are included in~\pn, an
immediate consequence of Remark~\ref{rk:adeq-clos} is the perfect
normalisation of typed \lcm-calculus.
\begin{thm}
  Every well typed term is perfectly normalising for \lcm.
\end{thm}

Furthermore, every closed and defined normal form is a value or the
Daimon (Proposition~\ref{prop:fn-val}).
Since the Daimon is never created by a reduction step, typing a term
ensures that it reduces strongly ---and without case composition--- on
a value.
We can even be more precise when using data types:
if a term (written without~$`?$) has type $\tco{c}T_1\dots T_k$, then
it reduces on a data structure $\co{c}t_1\dots t_k$
(Proposition~\ref{prop:val-dc}).

Now we call \emph{pure value} a data structure whose all sub-terms are
data structures (such as \co{cons 0 (cons (S(S0)) nil)} for instance)
and \emph{pure data type} a data type whose all sub-terms are data
types.

A pure value is trivially typable by a pure data type (just replace
every constructor~\co{c} in the term by the corresponding type
constructor~\tco{c} to obtain the type, and use \trul{Constr} and
\strul{\textstyle Data} to derive the typing judgement).
Conversely, every closed defined normal term without~$`?$ in a pure
data type is a pure value (by induction on the structure of the term,
using Proposition~\ref{prop:val-dc}).

Hence, if~$t$ is a term written without the Daimon, and~$D$ is a pure
data type,
\begin{displaymath}
  \these t:D \qquad\implies\qquad t 
  \text{ reduces strongly in \lcm\ on a pure value of } D
\end{displaymath}
(where a pure value of $\tco{c}D_1\dots D_k$ has form $\co{c}v_1\dots
v_k$ with $v_i$ a pure value of~$D_i$).

In that sense, we can say that case composition is unessential  in
this calculus:
it is not necessary to reach pure values.

\section*{Conclusion}

Typed lambda calculus with constructors provides a powerful
polymorphic type system, with a notion of data types and type
application.
The difficulty of typing the commutation rule between case and
application is overcome with a sub-typing system.
In this paper we have shown that this type system ensures strong
normalisation without match failure if \emph{we remove the composition
of case analysers} from the calculus.
We can safely do so, since the case composition rule is not
computationally necessary.
However, we thus lose the separation property for the lambda calculus
with constructors.

\paragraph*{Related works. }
The first presentation of the pattern calculus~~\cite{Jay04} comes
with a ML-style type system.
This type system is less expressive than ours and does not prevent match
failure during reduction, but it is decidable.

A more elaborated calculus, the \emph{extension calculus}, was recently
developed in~\cite{JayBook}.
It is typed with an extension of \textit{System~F \`a la Church}, 
that provides type application and also a pattern matching mechanism on
types.
A proof of strong normalisation, using the method based on
reducibility candidates, is done for a restriction of this system.
Although no type inference algorithm exists for this calculus, it has
been implemented in \texttt{bondi}~\cite{bondi}.

Several Church-style type systems have been proposed for the $`r$-calculus,
including a family of type systems organised in a cube similar to
Barendregt's.
As far as we know, no Curry-style type system has been proposed for the
$`r$-calculus.

\paragraph*{Future works. }
This paper has raised many questions, mainly concerning a possible
implementation of lambda calculus with constructors.
The first one is about recursively defined data types, such as
\begin{displaymath}
  \nat \equiv \mathtt{0} \cup \mathtt{succ}(\nat)
  \qquad;\qquad
  list\ T \equiv \mathtt{nil}\; \cup \; \mathtt{cons}\ T\ (list\ T)
\end{displaymath}
Adding a double sub-typing judgement for each data type is a way to do
it, but it requires checking the correctness of each rule.
A fixpoint operator would probably be a better way, since it would allow
to add recursive data types ``on the fly''.

Still with the view to implementing \lc-calculus, we need to isolate a
decidable fragment of our type system. This is a real challenge when
it comes to type case bindings (remind the example of
Section~\ref{sec:typ-cb} page~\pageref{page:indec}) and to use
union types.

Last, it could be interesting to develop a denotational semantic for
the lambda calculus with constructors.
Since the literature about denotational semantics for pure lambda
calculus (based on domain theory for instance) is abundant, 
we could try to adapt it to our calculus.
An idea to do that, is to first traduce \lc-calculus into pure
$`l$-calculus (in the spirit of CPS translations).
\subsubsection{Acknowledgements. }
I started this work at the University of Buenos Aires, which hosted me
for 6 months during my master thesis.
I would like to thank Ariel Arbiser, Eduardo Bonelli, Carlos Lombardi,
Alejandro R\'ios and Roel de Vrijer for all the discussions we had
there, and that were profitable for this paper.
I also acknowledge my supervisor, Alexandre Miquel, for his helpful
advice.

\bibliographystyle{plain}
\bibliography{Petit-tlcaSI10}

\end{document}